\documentclass[aps,12pt,endfloats*]{revtex4} 
\usepackage{graphicx}

\def\beq{\begin{equation}}
\def\eeq{\end{equation}}
\def\n{n}
\def\p{p}
\def\d{\delta}
\def\A{{\cal A}}
\def\B{{\cal B}}
\def\C{{\cal C}}

\def\H1{{H_1}}

\def\a00{{{\cal A}_0^0}}
\def\b00{{{\cal B}_0^0}}
\def\c00{{{\cal C}_0^0}}
\def\D00{{{\cal D}_0^0}}
\def\d{\delta}
\def\A{{\cal A}}
\def\B{{\cal B}}
\def\C{{\cal C}}

\def\mun{{\mu_n}}
\def\mup{{\mu_p}}

\begin{document}

%\voffset 0.5 true in

\title{Do Neutron Star Gravitational Waves Carry Superfluid Imprints?}

\author{G. L. Comer}

\affiliation{Department of Physics, Saint Louis University, 
St.~Louis, MO, 63156-0907, USA}

\begin{abstract}

\vskip 24pt

Isolated neutron stars undergoing non-radial oscillations are 
expected to emit gravitational waves in the kilohertz frequency 
range.  To date, radio astronomers have located about 1,300 
pulsars, and can estimate that there are about $2 \times 10^8$ 
neutron stars in the galaxy.  Many of these are surely old and 
cold enough that their interiors will contain matter in the 
superfluid or superconducting state.  In fact, the so-called 
glitch phenomenon in pulsars (a sudden spin-up of the pulsar's 
crust) is best described by assuming the presence of 
superfluid neutrons and superconducting protons in the inner 
crusts and cores of the pulsars.  Recently there has been much 
progress on modelling the dynamics of superfluid neutron stars 
in both the Newtonian and general relativistic regimes.  We will 
discuss some of the main results of this recent work, perhaps 
the most important being that superfluidity should affect the 
gravitational waves from neutron stars (emitted, for instance, 
during a glitch) by modifying both the rotational properties of 
the background star and the modes of oscillation of the perturbed 
configuration.  Finally, we present an analysis of the so-called 
zero-frequency subspace (i.e.~the space of time-independent 
perturbations) and determine that it is spanned by two sets of 
polar (or spheroidal) and two sets of axial (or toroidal) 
degenerate perturbations for the general relativistic system.  
As in the Newtonian case, the polar perturbations are the 
g-modes which are missing from the pulsation spectrum of a 
non-rotating configuration, and the axial perturbations should 
lead to two sets of r-modes when the degeneracy of the 
frequencies is broken by having the background rotate.     
\end{abstract}

\maketitle

\section{Introduction}

Jacob Bekenstein is one of those rare individuals who can make 
significant, original contributions to diverse areas of 
theoretical physics.  He is also a man of great integrity and, 
I believe, has a humility that serves him well in advising and 
supporting students and young scientists.  I am profoundly 
grateful that fate allowed me to be one of those young scientists 
and now lets me participate in this celebration of his career.  
One of the areas of theoretical physics that Jacob has worked on 
is relativistic fluid dynamics.  This is an important component 
of my current area of research, which is to develop models of 
Newtonian and general relativistic superfluid neutron stars.  My 
original interest in superfluids, appropriately enough, was 
sparked by Jacob, when he suggested that I look at superfluid 
analogs of effects predicted for quantum fields in curved 
spacetimes (the Hawking and Fulling-Davies-Unruh effects).  My 
current interest in superfluids is to determine how the dynamics 
of superfluid neutron stars differ from their ordinary, or 
perfect, fluid counterparts and if the different dynamics can 
lead to observable effects in gravitational waves.  In the 
remainder of this article, I will give an overview of what my  
collaborators and I have accomplished so far, including some new 
results (from work with Nils Andersson) on the structure of the 
so-called zero-frequency subspace (i.e.~the space of 
time-independent perturbations).  The main purpose is to show that 
superfluidity in neutron stars should affect their gravitational 
waves in two ways, by modifying the rotational properties of the 
background star and the modes of oscillation of the perturbed 
configuration.

While there are many mysteries about neutron stars that remain 
to be explained, we do have some significant observational facts 
to work with.  For instance, Lorimer \cite{L01} reports that 
nearly 1300 pulsars (i.e.~rotating neutron stars) have now been 
observed.   By extrapolating the data on the local population, 
he can estimate that there are about $1.6 \times 10^5$ normal 
pulsars and around $4 \times 10^4$ millisecond pulsars in our 
galaxy.  Of course, there are also neutron stars that are no 
longer active pulsars.  To get a handle on their number Lorimer 
takes the observed supernova rate, which is about 1 per 60 
years, and the age of the universe to find about $2 \times 
10^8$ neutron stars in the galaxy.  The overwhelming majority 
of these objects must be very cold in the sense that their 
(local) temperatures are much less than the (local) Fermi 
temperatures of the independent species of the matter.  One can 
estimate the Fermi temperature to be about $10^{12}~{\rm K}$ for 
neutrons at supra-nuclear densities, and it is generally accepted 
that within the first year (and probably much sooner than that) 
nascent neutron stars should cool to temperatures less than 
$10^9~{\rm K}$.  This is an interesting fact, in that nuclear 
physics calculations of the transition temperature for neutrons 
and protons to become superfluid and superconducting, 
respectively, consistently yield a value that is $10^9~{\rm K}$ 
in order of magnitude (for recent reviews see \cite{UL99,LS00}).  
Thus we can expect that a significant portion of the neutron 
stars in our galaxy will have at least two (and perhaps more) 
superfluids in their cores.

In addition to nuclear physics theory and experiment, the 
well-established glitch phenomenon in pulsars (e.g.~Vela and 
Crab) \cite{RM69,L93} is perhaps the best piece of evidence 
that supports the existence of superfluids in neutron stars.  
A glitch is a sudden spin-up of the observed rotation rate 
of a neutron star, and can have a relaxation time of weeks 
to months \cite{RD69}.  Baym et al \cite{BPPR69} have noted 
that a relaxation mechanism based on ordinary fluid viscosity 
would be much too short to explain a weeks to months timescale 
and so they argue that this signals the presence of a neutron 
superfluid.  Now, a mainstay idea for explaining glitches is 
that of superfluids and their vortex dynamics, i.e.~how the 
vortices get pinned, unpinned, and then repinned 
\cite{AI75,AAPS84a,AAPS84b} to nuclei in the inner crusts of 
the glitching pulsars.  This is known as the vortex creep 
model and in it glitches are a transfer of momentum via 
vortices from one angular momentum carrying component of the 
star to another.  The model has worked well to describe both 
the giant glitches in Vela and the smaller ones of the Crab.  
The vortex creep model can also be used to infer the internal 
temperature of a glitching pulsar, and for Vela it implies a 
temperature of $10^7~{\rm K}$ \cite{AAPS84b}.  It is also 
interesting to note the work of Tsakadze and Tsakedze 
\cite{TT80} who have experimented with rotating superfluid 
Helium II and find behaviour very much like glitches in 
pulsars.  

The classic description of superconductivity in ordinary 
condensed matter systems is based on the so-called ``BCS'' 
mechanism (see, for instance, \cite{TT86} or \cite{G85} 
for excellent presentations): the particles that become 
superconducting must be fermions, and below a certain 
transition temperature there must be an (usually 
effective) attractive interaction between them (at the 
Fermi surface with zero total momentum).  The interaction 
leads to so-called Cooper-pairing where a pair of fermions 
act like a single boson and a collection of them can behave 
as a condensate.  The mechanism is very robust, which is 
why it also forms the basis for discussion of nucleon 
superfluidity and superconductivity \cite{S89}; 
i.e.~nucleons are fermions and the effective interaction 
between them at nuclear and supra-nuclear densities can be 
attractive.  For instance, it is known experimentally that 
the lowest excited states in even-even nuclei are 
systematically higher than other nuclei because of pairing 
between nucleons which must be broken \cite{TT86}.  

After many years of development, beginning with the work 
of Migdal \cite{M59}, a consistent picture has emerged (in 
part, from gap calculations \cite{UL99,LS00}): At 
long-range the nuclear force is attractive and leads to 
neutron ``Cooper'' pairing in ${}^1{\rm S}_0$ states in the 
inner crust, but because of short-range repulsion in the 
nuclear force and the spin-orbit interaction neutrons pair 
into ${}^3{\rm P}_2$ states in the more dense regions of the 
core \cite{HGRR70}.  In the crust protons are locked inside 
of neutron rich nuclei embedded in a degenerate normal 
fluid of electrons.  In the inner crust the nuclei are also 
embedded in, and even penetrated by, the superfluid 
neutrons.  In the core, however, the nuclei have dissolved 
and the protons remain dilute enough that they feel only 
the long-range attractive part of the nuclear force and 
pair in ${}^1{\rm S}_0$ states.  There is no pairing 
between neutrons and protons anywhere in the core since 
their respective Fermi energies are so different.  The 
core superfluid neutrons and superconducting protons are 
also embedded in a highly degenerate normal fluid of 
electrons.  Other possibilities, such as pion or hyperon 
condenstates, have been put forward but we will keep to 
the simplest scenario that considers only superfluid 
neutrons, superconducting protons, crust nuclei, and normal 
fluid electrons.   

There are several ways in which the dynamics of a superfluid 
differ from its ordinary fluid counterpart, and each 
difference should have some impact on the gravitational 
waves that a superfluid neutron star emits.  One key 
difference is that a pure superfluid is locally irrotational.  
A superfluid, however, can mimic closely ordinary fluid 
rotation by forming a dense array of (quantized) vortices.  
In the core of each vortex the superfluidity is destroyed 
and the particles are in an ordinary fluid state, and can 
carry non-zero vorticity.  A second, very important difference 
is when there are several species of matter in a superfluid, 
or superconducting, state.  The superfluids of all the species 
will interpenetrate and each superfluid will be dynamically 
independent having its own unit four-vector and local particle 
number density.  Lastly, superfluids have zero viscosity, but 
when vortices and excitations are present, then dissipative 
mechanisms can exist.  For instance, the scattering of 
excitations off of the normal fluid in the vortex cores can 
lead to dissipative momentum exchange between the excitations 
and the superfluid, the net effect being that the superfluid 
motion becomes dissipative.  This form of dissipative 
mechanism is known as mutual friction.   

In neutron stars there is a very efficient form of mutual 
friction \cite{ALS84,AS88,S89} that depends on the entrainment 
effect \cite{AB75,VS81}, which Sauls \cite{S89} describes 
as follows: even though the neutrons are superfluid and the 
protons are superconducting both will still feel the long-range 
attractive component of the nuclear force.  In such a system of 
interacting fermions the resulting excitations are 
quasiparticles.  This means that the bare neutrons (or protons) 
are ``dressed'' by a polarization cloud of nucleons comprised 
of both neutrons and protons.  Since both types of nucleon 
contribute to the cloud the momentum of the neutrons, say, is 
modified so that it is a linear combination of the neutron and 
proton particle number density currents.  The same is true of 
the proton momentum.  Thus when one of the nucleon fluids 
starts to flow it will, through entrainment, induce a momentum 
in the other fluid.  Alpar et al \cite{ALS84} have shown that 
the electrons track very closely the superconducting protons 
(because of electromagnetic attraction).  Around each vortex is 
a flow of the superfluid neutrons.  Because of entrainment, a 
portion of the protons, and thus electrons too, will be 
pulled along with the superfluid neutrons.  The motion of the 
plasma leads to magnetic fields being attached to the vortices.  
The mutual friction in this case is the dissipative scattering 
of the normal fluid electrons off of the magnetic fields 
attached to the vortices.
    
There has been much effort put forward to develop Newtonian 
\cite{ML91,M91,LM94,LM95,P02} and general relativistic 
formalisms \cite{C85,C89,CL93,CL94,CL95,CL98a,CL98b,LSC98,CLS00} 
for describing superfluid neutron stars.  In the simplest, but 
still physically interesting, formalism one has a system that 
consists of two interpenetrating fluids---the superfluid 
neutrons in the inner crust and core and the remaining charged 
constituents (i.e.~crust nuclei, core protons, and crust and 
core electrons) that will be loosely referred to as 
``protons''---and the entrainment effect that acts between them.  
In principle the model can be expanded to have more than two 
interpenetrating fluids (see \cite{P00} for instance).  As well 
a given superfluid can be confined to a distinct region in the 
star \cite{ACL02}.  In this way the proton fluid, say, can be 
made to extend out farther than the superfluid neutrons.  This 
is a first approximation at incorporating the fact that the 
superfluid neutrons do not extend all the way to the surface 
of the star. 

Our primary goal is to show that superfluidity will affect 
gravitational wave emission from neutron stars.  We will 
see that a suitably advanced, but plausible, detector will 
have enough sensitivity at high frequency to see modes 
excited during a glitch.  With such detections we will be 
able to place constraints, say, on the parameters that 
describe entrainment.  But in addition to studying glitches, 
we also need to analyze further the recently discovered 
instability in the r-modes of neutron stars \cite{A98,FM98}.   
The instability is driven by gravitational wave emission (the 
CFS mechanism \cite{C70,FS78,F78}) and the waves are 
potentially detectable by LIGO II \cite{LOM98,OLCSVA98,AKS99a}.  
The conventional wisdom early on stated that mutual friction 
would act against the instability in a superfluid neutron star 
and thus effectively suppress the gravitational radiation.  
But Lindblom and Mendell \cite{LM00} have found that 
mutual friction is largely ineffective at suppressing the 
r-mode instability.  However, there are many questions 
about the spectrum of oscillation modes allowed by a rotating 
superfluid neutron star and the analysis of instabilities is 
very likely to be much richer than the ordinary fluid case.  
Thus, another goal here is to lay some groundwork for a 
future detailed study of the CFS mechanism in superfluid 
neutron stars.  Specifically, we will demonstrate that the 
zero frequency subspace is spanned by two sets of  
polar (or spheroidal) and two sets of axial (or toroidal) 
degenerate perturbations for the general relativistic system.  
Like the Newtonian case \cite{AC01b}, the polar perturbations 
are the g-modes which are missing from the pulsation spectrum 
of a non-rotating configuration, and the axial perturbations 
should lead to two sets of r-modes when the degeneracy of the 
frequencies is broken by having the background rotate.  

Below we will alternate between discussions based on Newtonian 
gravity and those using general relativity.  Accuracy demands 
that a fully relativistic formalism be employed, however there 
are some questions of principle for which a Newtonian formalism 
can suffice.  For instance, in determining the number of different 
modes of oscillation that a superfluid neutron star can undergo 
it is much more tractable to use the Newtonian equations.  But, 
ulitmately there is the need for a general relativistic formalism.  
Newtonian gravity does not include gravitational waves and so one 
needs a fully relativistic formalism to get an accurate damping 
time of a mode of oscillation due to gravitational wave emission.  
Also, there is the well-known problem that Newtonian models do 
not produce reliable values for the mass and radius, in that, for 
a given central density, the predicted mass and radius in Newtonian 
models may differ considerably from those of general relativity.  
This is a crucial point since the superfluid phase transition in 
a neutron star is sensitive to density, as are the parameters 
relevant to entrainment, and the oscillation frequencies can depend 
sensitively on mass and radius.  

\section{General Relativistic and Newtonian Superfluid Formalisms} 
\label{sfform}

Based on the preceeding discussion, our formalism for modelling 
superfluid neutron stars must allow for two interpenetrating fluids 
(i.e.~the neutrons and ``protons'') and the entrainment effect.  
This must be the case whether we are working in the general 
relativistic regime or the Newtonian limit.  A general relativistic 
superfluid formalism has been developed by Carter and Langlois 
\cite{C85,C89,CL95,CL98a,CL98b} and their collaborators 
\cite{CL93,CL94,LSC98,CLS00}.  Here an action principle will be 
outlined that yields the equations of motion and the stress-energy 
tensor for this system.  Although a variational principle also exists 
for the Newtonian regime \cite{P02}, we will briefly indicate how to 
get the Newtonian equations by taking the appropriate limit of the 
general relativistic equations (see \cite{AC01b} for full details).  
In the same spirit, we will use a ``slow'' velocity approximation to 
motivate an expansion that can be used to incorporate existing 
models of the entrainment effect \cite{ACL02}.

\subsection{The General Relativistic Formalism}

In this subsection the equations of motion and stress-energy tensor 
for a two-component general relativistic superfluid are obtained 
from an action principle.  Specifically a so-called ``pull-back'' 
approach (for instance, see \cite{CL93,CL94}) is used to construct 
Lagrangian displacements of the number density four-currents that 
form the basis of the variations of the fluid variables in the action 
principle (they will also be used later in the analysis of the 
zero-frequency subspace).  It will generalize to the superfluid case 
some of the techniques used to analyze the CFS mechanism in ordinary 
fluid neutron stars \cite{FS78,F78}.  Finally, it can serve as a 
starting point for generalizing the Hamiltonian formalism developed 
by Comer and Langlois \cite{CL94} who limited their discussion to a 
superfluid in the Landau state \cite{TT86} (i.e.~purely irrotational).  

For both the rotation and mode calculations effects such as 
``transfusion'' \cite{LSC98} of one component into the other (because 
of the weak interaction, for instance) will be ignored.  The neutron 
and proton number density four-currents, to be denoted $\n^{\mu}$ 
and $\p^{\mu}$ respectively, are thus taken to be separately conserved, 
meaning   
\beq
    \nabla_{\mu} \n^{\mu} = 0 \quad , \quad  
    \nabla_{\mu} \p^{\mu} = 0 \ . \label{consv}
\eeq
Such an approximation is reasonable for the mode calculations since 
the time-scale of the oscillations (which is milliseconds) is much 
less than the weak interaction time-scale in neutron stars, as long 
as the amplitudes of the oscillations remain small enough \cite{E88}.  
In the case of slowly rotating neutron stars, it has been shown 
\cite{AC01a,PCA02} that when the neutrons and protons rotate rigidly 
at different rates then chemical equilibrium cannot exist between 
them.  Of course, the energy associated with the relative rotation 
could be dissipated through a process like transfusion.  But Haensel 
\cite{H92} has demonstrated that such a process would take months to 
years in mature neutron stars, and so again it will be neglected 
(since ultimately we are interested in gravitational waves emitted 
during a glitch, say, the timescale of which would be much shorter 
than the transfusion timescale).

By introducing the duals to $\n^{\mu}$ and $\p^{\mu}$, i.e. 
\beq
    \n_{\nu \lambda \tau} = \epsilon_{\nu \lambda \tau \mu} 
                            \n^{\mu} \quad , \quad
    \n^{\mu} = { 1 \over 3!} \epsilon^{\mu \nu \lambda \tau} 
               \n_{\nu \lambda \tau} \ ,
\eeq
and
\beq
    \p_{\nu \lambda \tau} = \epsilon_{\nu \lambda \tau \mu} 
                            \p^{\mu} \quad , \quad
    \p^{\mu} = { 1 \over 3!} \epsilon^{\mu \nu \lambda \tau} 
               \p_{\nu \lambda \tau} \ ,
\eeq
respectively, then the conservation rules are equivalent to 
having the two three-forms be closed, i.e.
\beq
    \nabla_{[\mu} \n_{\nu \lambda \tau]} = 0 \quad , \quad
    \nabla_{[\mu} \p_{\nu \lambda \tau]} = 0 \ .
\eeq
The reason for introducing the duals is that it is 
straightforward to construct particle number density three-forms 
that are automatically closed.  The point is that the 
conservation of the particle number density currents should 
not---speaking from a strict field theoretic point of view---be 
a part of the system of equations of motion, rather they should 
be automatically satisfied when the ``true'' system of equations 
is imposed.  

This can be made to happen by introducing two three-dimensional 
abstract spaces which can be labelled by coordinates $X^A$ and 
$Y^A$, respectively, where $A,B,C~{\rm etc} = 1,2,3$.  By 
``pulling-back'' each  three-form onto its respective abstract 
space we can construct three-forms that are automatically 
closed on spacetime, i.e.~let
\beq
    \n_{\nu \lambda \tau} = N_{A B C}(X^D) \nabla_\nu X^A 
                            \nabla_\lambda X^B \nabla_\tau 
                            X^C \quad , \quad
    \p_{\nu \lambda \tau} = P_{A B C}(Y^D) \nabla_\nu Y^A 
                            \nabla_\lambda Y^B \nabla_\tau 
                            Y^C
\eeq
where $N_{A B C}$ and $P_{A B C}$ are completely antisymmetric 
in their indices.  Because the abstract space indices are 
three-dimensional and the closure condition involves four 
spacetime indices, and also that the $X^A$ and $Y^A$ are scalars 
on spacetime (and thus two covariant differentiations commute), 
the pull-back construction does indeed produce a closed three-form:  
\beq
   \nabla_{[\mu} \n_{\nu \lambda \tau]} = \nabla_{[\mu} \left(
           N_{A B C}(X^D) \nabla_\nu X^A \nabla_\lambda X^B 
           \nabla_{\tau]} X^C\right) \equiv 0 \ ,  
\eeq
and similarly for the protons.  In terms of the scalar fields 
$X^A$ and $Y^A$, we now have particle number density currents that 
are automatically conserved, and so another way of viewing the 
pull-back construction is that the fundamental fluid field 
variables are the spacetime functions $X^A$ and $Y^A$ 
\cite{rationale}.  The variations of the three-forms can now be 
derived by varying them with respect to $X^A$ and $Y^A$.  

Let us introduce two Lagrangian displacements on spacetime for the 
neutrons and protons, to be denoted $\xi^\mu_\n$ and $\xi^\mu_\p$, 
respectively.  These are related to the variations $\delta X^A$ and 
$\delta Y^A$ via a  ``push-forward'' construction:
\beq
    \delta X^A = - \left(\nabla_\mu X^A\right) \xi^\mu_\n 
                 \quad , \quad 
    \delta Y^A = - \left(\nabla_\mu Y^A\right) \xi^\mu_\p \ .
\eeq
Using the fact that
\begin{eqnarray}
    \nabla_\nu \delta X^A &=& - \nabla_\nu \left(\left[\nabla_\mu 
                              X^A\right] \xi^\mu_\n\right) \cr
                           && \cr
                          &=& - \left(\nabla_\mu X^A\right) 
                              \nabla_\nu \xi^\mu_\n - \left(
                              \nabla_\mu \nabla_\nu X^A\right) 
                              \xi^\mu_\n \ ,
\end{eqnarray}
and similarly for the proton variation, then we find \cite{LSC98}  
\begin{eqnarray}
    \delta \n_{\nu \lambda \tau} &=& - \left(\xi^\sigma_\n 
           \nabla_\sigma \n_{\nu \lambda \tau} + \n_{\sigma 
           \lambda \tau} \nabla_\nu \xi^\sigma_\n + 
           \n_{\nu \sigma \tau} \nabla_\lambda \xi^\sigma_\n
           + \n_{\nu \lambda \sigma} \nabla_\tau \xi^\sigma_\n
           \right)  
           = - {\cal L}_{\xi_\n} \n_{\nu \lambda \tau} \ , \cr
           && \cr          
    \delta \p_{\nu \lambda \tau} &=& - \left(\xi^\sigma_\p 
           \nabla_\sigma \p_{\nu \lambda \tau} + \p_{\sigma 
           \lambda \tau} \nabla_\nu \xi^\sigma_\p + 
           \p_{\nu \sigma \tau} \nabla_\lambda \xi^\sigma_\p
           + \p_{\nu \lambda \sigma} \nabla_\tau \xi^\sigma_\p
           \right)  
           = - {\cal L}_{\xi_\p} \p_{\nu \lambda \tau} \ , 
\end{eqnarray}
where ${\cal L}$ is the Lie derivative.  We can thus infer that
\begin{eqnarray}
    \delta n^\mu &=& \n^\sigma \nabla_\sigma \xi^\mu_\n - 
                     \xi^\sigma_\n \nabla_\sigma \n^\mu - \n^\mu 
                     \left(\nabla_\sigma \xi^\sigma_\n + 
                     {1 \over 2} g^{\sigma \rho} \delta 
                     g_{\sigma \rho}\right) \quad , \quad  \cr
                  && \cr
    \delta p^\mu &=& \p^\sigma \nabla_\sigma \xi^\mu_\p -
                     \xi^\sigma_\p \nabla_\sigma \p^\mu - \p^\mu 
                     \left(\nabla_\sigma \xi^\sigma_\p + 
                     {1 \over 2} g^{\sigma \rho} \delta 
                     g_{\sigma \rho}\right) \ .
\end{eqnarray}
By introducing the two decompositions
\begin{eqnarray}
    \n^\mu &=& n u^\mu \quad , \quad u_\mu u^\mu = - 1 \ , \cr
            && \cr
    \p^\mu &=& p v^\mu \quad , \quad v_\mu v^\mu = - 1 \ ,
\end{eqnarray}
then we can show furthermore that
\begin{eqnarray}
   \delta \n &=& - \nabla_\sigma\left(\n \xi^\sigma_\n\right) - 
                 \n \left(u_\nu u^\sigma \nabla_\sigma \xi^\nu_\n 
                 + {1 \over 2} \left[g^{\sigma \rho} + u^\sigma 
                 u^\rho\right] \delta g_{\sigma \rho}\right)  
                 \ , \cr
              && \cr
   \delta \p &=& - \nabla_\sigma\left(\p \xi^\sigma_\p\right) - 
                 \p \left(v_\nu v^\sigma \nabla_\sigma \xi^\nu_\p 
                 + {1 \over 2} \left[g^{\sigma \rho} + v^\sigma 
                 v^\rho\right] \delta g_{\sigma \rho}\right) 
                 \ , \label{dens_perbs}                 
\end{eqnarray}
and
\begin{eqnarray}
   \delta u^\mu &=& \left(\delta^\mu_\rho + u^\mu u_\rho\right) 
                    \left(u^\sigma \nabla_\sigma \xi^\rho_\n  - 
                    \xi^\sigma_\n \nabla_\sigma u^\rho\right) + 
                    {1 \over 2} u^\mu u^\sigma u^\rho 
                    \delta g_{\sigma \rho} \ , \cr
      && \cr
   \delta v^\mu &=& \left(\delta^\mu_\rho + v^\mu v_\rho\right) 
                    \left(v^\sigma \nabla_\sigma \xi^\rho_\p - 
                    \xi^\sigma_\p \nabla_\sigma v^\rho\right) + 
                    {1 \over 2} v^\mu v^\sigma v^\rho 
                    \delta g_{\sigma \rho} \ . \label{vel_perbs}
\end{eqnarray}

We will take one more step ahead and associate a notion of Lagrangian 
variation with each Lagrangian displacement.  These are defined to 
be
\begin{equation}
    \Delta_\n \equiv \delta + {\cal L}_{\xi_\n} \quad , \quad 
    \Delta_\p \equiv \delta + {\cal L}_{\xi_\p} \ , 
\end{equation}
so that it now follows that 
\beq
    \Delta_\n \n_{\mu \lambda \tau} = 0 \quad , \quad 
    \Delta_\p \p_{\mu \lambda \tau} = 0 \ ,
\eeq
which is entirely consistent with the pull-back construction.  We 
also find that
\beq
    \Delta_\n u^\mu = {1 \over 2} u^\mu u^\sigma u^\rho \Delta_\n 
                      g_{\sigma \rho} \quad , \quad 
    \Delta_\p v^\mu = {1 \over 2} v^\mu v^\sigma v^\rho \Delta_\p 
                      g_{\sigma \rho} \ ,
\eeq 
\beq
    \Delta_\n \epsilon_{\nu \lambda \tau \sigma} = {1 \over 2} 
                      \epsilon_{\nu \lambda \tau \sigma} 
                      g^{\mu \rho} \Delta_\n 
                      g_{\mu \rho} \quad , \quad 
    \Delta_\p \epsilon_{\nu \lambda \tau \sigma} = {1 \over 2} 
                      \epsilon_{\nu \lambda \tau \sigma} 
                      g^{\mu \rho} \Delta_\p 
                      g_{\mu \rho} \ ,
\eeq
and
\beq
    \Delta_\n \n = - {\n \over 2} \left(g^{\sigma \rho} + 
                   u^\sigma u^\rho\right) \Delta_\n g_{\sigma 
                   \rho} 
                   \quad , \quad 
    \Delta_\p \p = - {\p \over 2} \left(g^{\sigma \rho} + 
                   v^\sigma v^\rho\right) \Delta_\p g_{\sigma 
                   \rho} \ .
\eeq
However, in contrast to the ordinary fluid case \cite{FS78,F78}, 
there are many more options to consider.  For instance, we could 
also look at the Lagrangian variation of the neutron number density 
with respect to the proton flow, i.e.~$\Delta_\p \n $, or the 
Lagrangian variation of the proton number density with respect to 
the neutron flow, i.e.~$\Delta_\n \p$.  It is not clear at this 
point how the existence of two preferred rest frames---one that is 
attached to the neutrons and the other that is attached to the 
protons---will affect an analysis of the CFS mechanism in superfluid 
neutron stars.  

Nevertheless, with a general variation of the conserved four-currents 
in hand, we can now use an action principle to derive the equations 
of motion and the stress-energy tensor.  The central quantity is the 
so-called ``master'' function $\Lambda$, which is a function of all 
the different scalars that can be formed from $\n^{\mu}$ and $\p^{\mu}$, 
i.e.~the three scalars $\n^2 = - \n_{\mu} \n^{\mu}$, $\p^2 = - \p_{\mu} 
\p^{\mu}$, and $x^2 = - \p_{\mu} \n^{\mu}$.  In the limit where the two 
currents are parallel, i.e.~the two fluids are comoving, then the master 
function is such that $- \Lambda$ corresponds to the local thermodynamic 
energy density.  In the action principle, the master function is the 
Lagrangian density for the two fluids.

An unconstrained variation of $\Lambda(\n^2,\p^2,x^2)$ with respect 
to the independent vectors $\n^{\mu}$ and $\p^{\mu}$ and the metric 
$g_{\mu \nu}$ takes the form
\beq
     \d \Lambda = \mu_{\mu} \d \n^{\mu} + \chi_{\mu} \d \p^{\mu} + 
                  {1 \over 2} \left(\n^{\mu} \mu^{\nu} + \p^{\mu} 
                  \chi^{\nu}\right) \d g_{\mu \nu}\ ,
\eeq
where 
\beq
     \mu_{\mu} = \B \n_{\mu} + \A \p_{\mu} \quad , \quad
     \chi_{\mu} = \C \p_{\mu} + \A \n_{\mu} \ , \label{muchidef}
\eeq
and
\beq
   \A = - {\partial \Lambda \over \partial x^2} \qquad , \qquad \B = 
        - 2 {\partial \Lambda \over \partial \n^2} \qquad , \qquad
   \C = - 2 {\partial \Lambda \over \partial \p^2} \ . \label{coef1}
\eeq
The momentum covectors $\mu_{\mu}$ and $\chi_{\mu}$ are dynamically, 
and thermodynamically, conjugate to $\n^{\mu}$ and $\p^{\mu}$, and 
their magnitudes are, respectively, the chemical potentials of the 
neutrons and the protons.  The two momentum covectors also show the 
entrainment effect since it is seen explicitly that the momentum of 
one constituent carries along some mass current of the other 
constituent (for example, $\mu_{\mu}$ is a linear combination of 
$\n^{\mu}$ and $\p^{\mu}$).  We also see that entrainment vanishes 
if $\Lambda$ is independent of $x^2$ (because then ${\cal A}=0$).

If the variations of the four-currents were left unconstrained,  
the equations of motion for the fluid implied from the above variation 
of $\Lambda$ would require, incorrectly, that the momentum covectors 
should vanish in all cases.  This reflects the fact that the variations 
of the conserved four-currents must be constrained.  In terms of the 
constrained Lagrangian displacements, a variation of $\Lambda$ now 
yields
\beq
    \d \left(\sqrt{- g} \Lambda\right) = {1 \over 2} \sqrt{- g} 
    \left(\Psi g^{\mu \nu} + \p^{\mu} \chi^{\nu} + \n^{\mu} 
    \mu^{\nu}\right) \d g_{\mu \nu} - 2 \sqrt{- g} \left(\n^{\mu} 
    \nabla_{[\mu} \mu_{\nu]} \xi^{\nu}_{\n} + \p^{\mu} \nabla_{[\mu} 
    \chi_{\nu]} \xi^{\nu}_{\p}\right) + T.D. 
\eeq
where the $T.D.$~is a total divergence and thus does not contribute 
to the field equations or stress-energy tensor.  At this point we 
can return to the view that $\n^{\mu}$ and $\p^{\mu}$ are the 
fundamental variables for the fluids.  Thus the equations of motion 
consist of the two original conservation conditions of 
Eq.~(\ref{consv}) plus two Euler type equations 
\beq
     \n^{\mu} \nabla_{[\mu} \mu_{\nu]} = 0 \qquad , \qquad \p^{\mu} 
          \nabla_{[\mu} \chi_{\nu]} = 0 \ , \label{eueqn} 
\eeq
since the two Lagrangian displacements are independent.  We see that 
the stress-energy tensor is  
\beq
     T^{\mu}_{\nu} = \Psi \delta^{\mu}_{\nu} + \p^{\mu} \chi_{\nu} 
                   + \n^{\mu} \mu_{\nu} \ , \label{seten} 
\eeq
where the generalized pressure $\Psi$ is defined to be  
\beq
     \Psi = \Lambda - \n^{\mu} \mu_{\mu} - \p^{\mu} \chi_{\mu} \ .
\eeq
When the complete set of field equations is satisfied then it is 
automatically true that $\nabla_{\mu} T^{\mu}_{\nu} = 0$. 

In a later section we will be interested in the linearized version 
of the combined Einstein and superfluid equations.  It will thus be 
convenient to write down the variations of the momentum covectors 
in terms of the variations of the particle number density currents.   
Following the scheme of Carter \cite{C89}, and Comer et al 
\cite{CLL99}, the variations of $\mu_{\mu}$ and $\chi_{\mu}$ due to 
a generic variation of $\n^{\mu}$, $\p^{\mu}$, and $g_{\mu \nu}$ take 
the form
\beq
    \d \mu_\rho = \A_\rho^{\ \sigma} \d \p_\sigma + \B_\rho^{\  \sigma}
                  \d \n_\sigma + \left(\d_g \A\right) \p_\rho + 
                  \left(\d_g \B\right) \n_\rho \ , 
\eeq
\beq
     \d \chi_\rho = \C_\rho^{\ \sigma} \d \p_\sigma + \A^\sigma_{\ 
                    \rho} \d \n_\sigma + \left(\d_g \C\right) \p_\rho + 
                    \left(\d_g \A\right) \n_\rho \ , \label{mom_perbs}
\eeq
with
\begin{eqnarray}
\A_{\mu \nu} &=& \A g_{\mu \nu} - 2 {\partial \B \over \partial \p^2} 
              \n_\mu \p_\nu - 2{\partial \A \over \partial \n^2} \n_\mu 
              \n_\nu - 2 {\partial \A \over \partial \p^2} \p_\mu \p_\nu 
              - {\partial \A \over \partial x^2} \p_\mu \n_\nu \ , \cr
              && \cr
\B_{\mu \nu} &=& \B g_{\mu \nu} - 2 {\partial \B \over \partial \n^2} 
              \n_\mu \n_\nu - 4 {\partial \A \over \partial \n^2} 
              \p_{(\mu} \n_{\nu)} - {\partial \A \over \partial x^2} 
              \p_\mu \p_\nu \ , \cr
              && \cr
\C_{\mu \nu} &=& \C g_{\mu \nu} - 2 {\partial \C \over \partial \p^2} 
              \p_\mu \p_\nu - 4 {\partial \A \over \partial \p^2} 
              \p_{(\mu} \n_{\nu)} - {\partial \A \over \partial x^2} 
              \n_\mu \n_\nu \ , \label{coefficients}
\end{eqnarray}
and the terms $\d_g \A$, $\d_g \B$ and $\d_g \C$ are given by 
\beq
\d_g \A = \left[{\partial \A \over \partial \n^2} \n^\mu \n^\nu
          + {\partial \A \over \partial \p^2} \p^\mu \p^\nu + {\partial 
          \A \over \partial x^2} \n^\mu \p^\nu\right] \d g_{\mu\nu} \ 
\label{dgA}
\eeq
($\d_g \B$ and $\d_g \C$ being given by analogous formulas, with $\A$ 
replaced by $\B$ and $\C$ respectively).  
 
\subsection{The Newtonian Limit}

The Newtonian superfluid equations can be obtained by writing the 
general relativistic field equations to order $c^0$, where $c$ is 
the speed of light, and then taking the limit that $c$ becomes 
infinite.  To order $c^0$ the metric can be written as
\beq
    ds^2 = - c^2 \left(1 + {2 \Phi \over c^2}\right)  d t^2 + 
           \d_{i j} d x^i  d x^j \ ,
\eeq
where the $x^i$ ($i = 1,2,3$) are Cartesian-like coordinates, and 
the gravitational potential $\Phi$ is assumed to be small in the 
sense that $- 1 << \Phi/c^2 \leq 0$.  To the same order the unit 
four-velocity components defined earlier are given by
\beq
    u^t = 1 - {\Phi \over c^2} + {v_{\n}^2 \over 2 c^2} 
          \ , \ 
    u^i = v_{\n}^i \ ,
\eeq
and
\beq
    v^t = 1 - {\Phi \over c^2} + {v_{\p}^2 \over 2 c^2} 
          \ , \ 
    v^i = v_{\p}^i \ ,
\eeq
where $v_{\n,\p}^2 = \d_{ij}v_{\n,\p}^i v_{\n,\p}^j$, and $v_{\n}^i$ 
and $v_{\p}^i$ are the Newtonian three-velocities of the neutron and 
proton fluids, respectively.  The three-velocities are assumed to be 
small with respect to the speed of light.  The entrainment variable 
$x^2$ takes the limiting form
\beq
    x^2 = \n_\n \n_\p \left(1 + {w^2 \over 2 c^2}\right) \ , 
          \label{xlimit}
\eeq
where 
\beq
    w^2 = \d_{ij} \left(v_{\n}^i - v_{\p}^i\right) 
                  \left(v_{\n}^j - v_{\p}^j\right) 
\eeq
and we have introduced the new notation $\n_\n = \n$ and $n_\p 
= \p$ (to distinguish between Newtonian quantities and their 
general relativity counterparts).  Finally, we separate the 
``master'' function $\Lambda$ into its mass part and a much 
smaller internal energy part $E$, i.e.~we write it as
\beq
   \Lambda = - \left(m_{\n} \n_\n + m_{\p} \n_\p\right) c^2 - 
               E(\n^2_\n,\n^2_\p,x^2) \ ,
\eeq
where $m_{\n}$ ($m_{\p}$) is the neutron (proton) mass.  

To more closely agree with the Newtonian superfluid equations 
derived by other means \cite{P02}, a different choice for the 
independent variables is used which is the triplet of variables 
$(\n^2_\n,\n^2_\p,w^2)$.  In this case $E = 
E(\n^2_\n,\n^2_\p,w^2)$ and the analog of the combined First 
and Second Law of Thermodynamics for the system takes the form
\beq
    d E = \mun d \n_\n + \mup d \n_\p + \alpha dw^2 \ ,
\eeq
where
\beq
    \mun = {\partial E \over \partial \n_\n} 
           \quad , \quad 
    \mup = {\partial E \over \partial \n_\p} 
           \quad , \quad 
    \alpha = {\partial E \over \partial w^2} \ . 
\eeq
The generalized pressure $\Psi$, to be renamed $P$, takes the 
limiting form
\beq
    P = - E + \mun \n_{\n} + \mup \n_{\p} \ .
\eeq

Formally letting the speed of light become infinite in the 
combined Einstein and general relativistic superfluid field 
equations, results in the following set of 9 equations:
\begin{eqnarray}
   0 &=& {\partial \n_\n \over \partial t} + \partial_i \left(\n_\n 
       v_{\n}^i\right) \ , \cr 
      && \cr
   0 &=& {\partial \n_\p \over \partial t} + \partial_i \left(\n_\p 
       v_{\p}^i\right) \ , \label{fullconsv}
\end{eqnarray}
and
\begin{eqnarray}
   0 &=& {\partial \over \partial t} \left(v_{\n}^i + {2 \alpha \over 
         m_{\n} \n_\n} \left[v_{\p}^i - v_{\n}^i\right]\right) + 
         v_{\n}^j \partial_j \left(v_{\n}^i + {2 \alpha \over m_{\n} 
         \n_\n} \left[v_{\p}^i - v_{\n}^i\right]\right) + \d^{ij} 
         \partial_j \left(\Phi + {\mu_\n \over m_{\n}}\right) + \cr
      && \cr
      &&{2 \alpha \over m_{\n} \n_\n} \d^{ij} \d_{kl} \left(v_{\p}^l - 
         v_{\n}^l\right) \partial_j v_{\n}^k \ , \cr
      && \cr
   0 &=& {\partial \over \partial t} \left(v_{\p}^i + {2 \alpha \over 
         m_{\p} \n_\p} \left[v_{\n}^i - v_{\p}^i\right]\right) + 
         v_{\p}^j \partial_j \left(v_{\p}^i + {2 \alpha \over m_{\p} 
         \n_\p} \left[v_{\n}^i - v_{\p}^i\right]\right) + \d^{ij} 
         \partial_j \left(\Phi + {\mu_\p \over m_{\p}}\right) + \cr
      && \cr 
      &&{2 \alpha \over m_{\p} \n_\p} \d^{ij} \d_{kl} \left(v_{\n}^l - 
         v_{\p}^l\right) \partial_j v_{\p}^k  \ . \label{fulleuler}
\end{eqnarray}
The gravitational potential $\Phi$ is obtained from 
\beq
    \partial_i \partial^i \Phi = 4 \pi G \left(m_\n \n_\n + m_\p \n_\p
                                 \right) \ . \label{poisson}
\eeq
For this system having no entrainment means setting the coefficient 
$\alpha$ to zero.  

These equations are equivalent to those derived independently by Prix 
\cite{P02}, using a Newtonian variational principle.  We also note 
that they are formally equivalent to the two-fluid set developed by 
Landau \cite{P74} for superfluid He~II.  The two fluids in the Landau 
case are traditionally taken to be the normal fluid (i.e.~the phonons, 
rotons, and other excitations) that carries the entropy and the rest 
of the fluid (i.e.~the superfluid) that carries no entropy.

\subsection{An analytical equation of state with entrainment}

The item that connects the microphysics to the global structure and 
dynamics of superfluid neutron stars is the master function, since 
it incorporates all of the information about the local thermodynamic 
state of the matter.  Ultimately, realistic models of superfluid 
neutron stars must be built using realistic master functions 
(i.e.~equations of state).  Only in this way can gravitational wave 
data be used to greatest effect to constrain the microphysics, such 
as parameters that are important for entrainment.  Unfortunately, 
there is not yet a fully relativistic determination of entrainment 
(although Comer and Joynt are currently working on this), and the 
best that can be done is to adapt models that have been used in the 
Netwonian limit.  Here we will describe an analytic expansion of the 
master function that facilitates the process (see \cite{ACL02} for 
all the details).    

In the limit where the fluid velocities are small with respect to 
$c$ we see from Eq.~(\ref{xlimit}) that the combination $x^2 - \n \p$ 
is small.  Thus, it makes sense to consider an expansion of the master 
function of the form \cite{ACL02}
\beq
    \Lambda(n^2,p^2,x^2) = \sum_{i = 0}^{\infty} \lambda_i(n^2,p^2) 
                           \left(x^2 - \n \p\right)^i \ .
\eeq
The $\A$, $\B$, and $\C$ coefficients that appear in the definitions 
of the momentum covectors become 
\begin{eqnarray}
  \A &=& - \sum_{i = 1}^{\infty} i~\lambda_i(n^2,p^2) \left(x^2 - \n 
         \p\right)^{i - 1}  \ , \cr
     && \cr
  \B &=& - {1 \over \n} {\partial \lambda_0 \over \partial \n} - {\p 
         \over \n} \A - {1 \over \n} \sum_{i = 1}^{\infty} {\partial 
         \lambda_i \over \partial \n} \left(x^2 - \n \p\right)^i \ , 
         \cr
     && \cr
  \C &=& - {1 \over \p} {\partial \lambda_0 \over \partial \p} - {\n 
         \over \p} \A - {1 \over \p} \sum_{i = 1}^{\infty} {\partial 
         \lambda_i \over \partial \p} \left(x^2 - \n \p\right)^i \ .  
\end{eqnarray}
The expansion is especially useful for both the rotation and mode 
calculations, since $x^2 = \n \p$ for any zeroth-order, or 
background, quantity.  In practice this means that only the first 
few $\lambda_i$ contribute, and for the mode calculations we need 
retain only $\lambda_0$ and $\lambda_1$.  The first coefficient 
$\lambda_0$ is directly related to equations of state that describe 
a mixture of neutrons and protons that are locally at rest with 
respect to each other.  The other coefficient $\lambda_1$ contains 
the information concerning the entrainment effect.  

\begin{figure}[t]
\centering
\caption{}
\vskip 24pt
\includegraphics[height=10cm,clip]{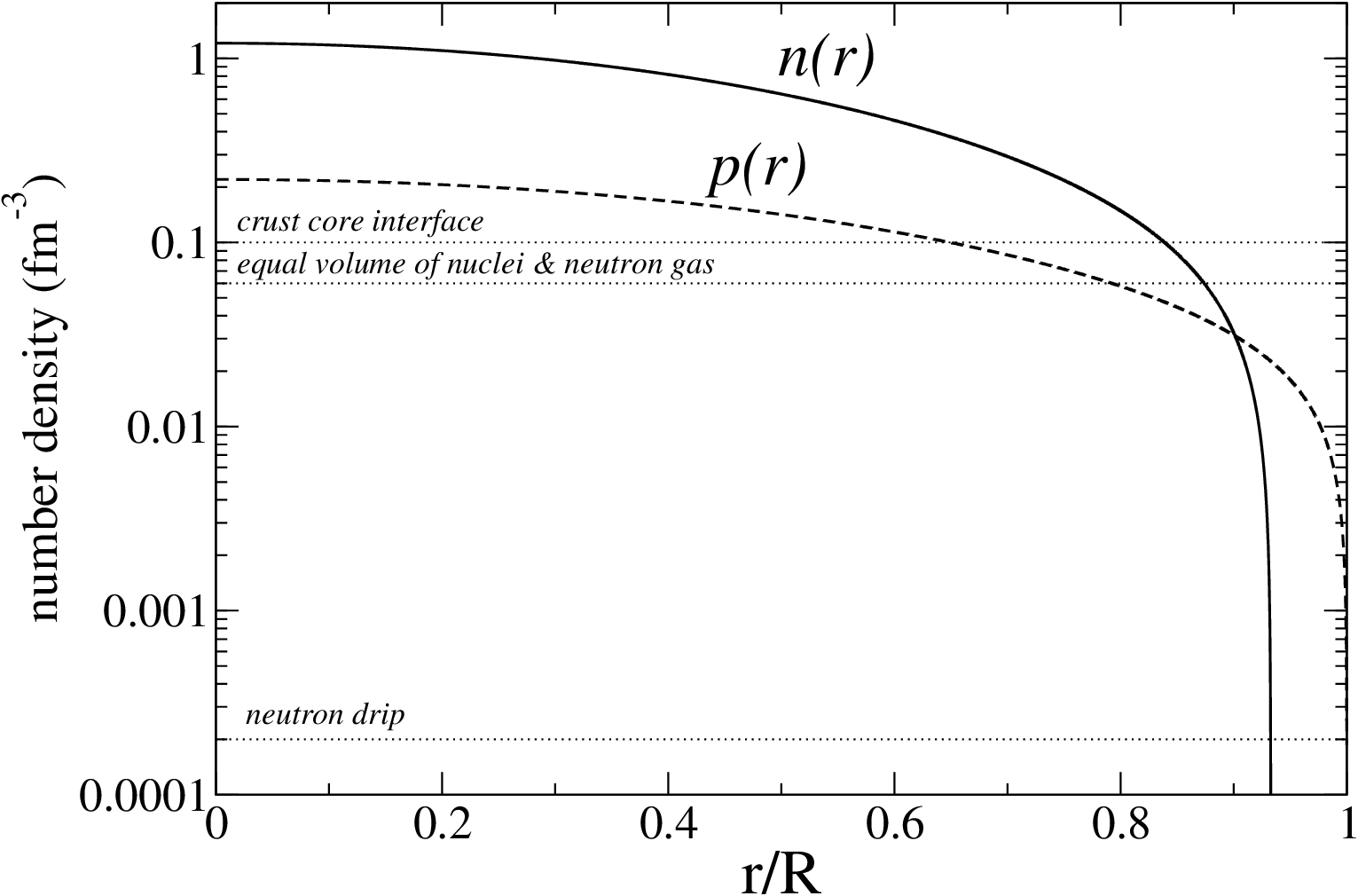}
\label{background}
\end{figure}

We will use a sum of two polytropes for $\lambda_0$, i.e.
\beq
    \lambda_0(\n^2,\p^2) = - m_\n \n - \sigma_{\n} \n^{\beta_{\n}}
                           - m_\p \p - \sigma_{\p} \p^{\beta_{\p}}
                              \ .
\eeq
In Table \ref{modtab} are given various parameter values that 
have been used in specific applications \cite{CLL99,AC01a,ACL02}.  
Model~I corresponds to the values used in the initial study of 
modes on a non-rotating background by Comer et al \cite{CLL99}, 
and is meant to describe only a neutron star core without an outer 
envelope of ordinary fluid.  Thus, the neutron and proton number 
densities vanish at the same radius.  Model~II is used in the mode 
study of Andersson et al \cite{ACL02} and is the most realistic.  
Its parameter values given in Table~\ref{modtab} have been chosen 
specifically to yield a canonical static and spherically symmetric 
background model having the following characteristics (see 
\cite{ACL02} for complete details): (i) A mass of about $1.4 
M_\odot$, (ii) a total radius of about $10\ {\rm km}$, (iii) an 
outer envelope of ordinary fluid of roughly one kilometer 
thickness, and (iv) a central proton fraction of about $10 \%$.  
These are values that are considered to be representative of a 
typical neutron star.  The background distribution of particles 
for this model is determined in Andersson et al \cite{ACL02} and 
their graph and caption is reproduced here in Fig.~\ref{background}.  
Finally, model~III is used by Andersson and Comer \cite{AC01a} to 
study slowly rotating configurations and is meant to be more 
realistic than model~I by extending the radius out further and 
having a better total mass value, but with no distinct envelope 
so that the neutron and proton number densities vanish at the same 
radius.  

Our strategy for incorporating entrainment is to adapt models that 
have been used in Newtonian calculations.  The particular model we 
will use is that of Lindblom and Mendell \cite{LM00}.  It is a 
parametrized approximation of the more detailed model based on 
Fermi liquid theory \cite{PN66} developed by Borumand, Joynt, and 
Klu\'zniak \cite{BJK96}.  As demonstrated in Andersson et al 
\cite{ACL02} the Lindblom and Mendell model translates into the 
relation    
\beq
    \lambda_1 = - {m_{\n} m_{\p} \over \rho_{\n\p}^2 - 
                \rho_{\n\n} \rho_{\p\p}} \rho_{\n\p} \ ,
\eeq
where
\beq
    m_{\n} \n = \rho_{\n \n} + \rho_{\n \p} \quad , \quad
    m_{\p} \p = \rho_{\p \p} + \rho_{\n \p} \ ,
\eeq
and
\beq
    \rho_{\n \p} = - \epsilon m_{\n} \n \ .
\eeq
Here $\epsilon$ is taken to be a constant, which is the approximation 
introduced by Lindblom and Mendell \cite{LM00}.  Prix, Comer and 
Andersson \cite{PCA02} argue that  
\beq
    \epsilon = {m_\p \p \over m_\n \n} \left({m_\p \over m^*_\p} - 
               1\right) \ ,
\eeq 
where $m^*_\p$ is the proton effective mass.  But, Sj\"oberg 
\cite{S76} has determined that $0.3 \leq m^*_\p/m_\p  \leq 0.8$.  
Assuming a proton fraction of about $10\%$ we can take as ``physical'' 
for a neutron star core those values that lie in the range $0.04 \leq 
\epsilon \leq 0.2$. 

\section{Slowly Rotating Superfluid Neutron Stars}

The first application to consider is to the problem of axisymmetric, 
stationary, and asymptotically flat configurations, i.e.~the 
standard assumptions made for rotating neutron stars 
\cite{C69,BC70,BGSM93}.  Although accurate codes exist for looking 
at rapidly rotating neutron stars in the ordinary fluid case 
\cite{BGSM93,GHLPBM99,S98}, and can in principle be adapted to the 
superfluid case (Prix, Novak, and Comer, work in progress), we will 
limit our discussion to situations where rapid rotation accuracy is 
not needed.  That is, we will describe models of rotating superfluid 
neutron stars that have been developed \cite{P99,AC01a,PCA02} using 
a slow-rotation approximation.  In the general relativistic regime, 
we discuss results of Andersson and Comer \cite{AC01a} who have 
adapted the one-fluid formalism of Hartle \cite{H67} and Hartle and 
Thorne \cite{HT68} to the superfluid case, whereas in the Newtonian 
limit it is the work Prix et al \cite{PCA02} (who have built on 
the Chandrasekhar-Milne approach \cite{C33,M23}) that will be 
reviewed.  The most important aspect of the superfluid case in 
both the general relativistic regime and Newtonian limit is that 
the neutrons can rotate at a rate different from that of the 
protons, and this obviously has no analog in the ordinary fluid 
case.

At the heart of the slow-rotation approximation is the assumption 
that the star is rotating slowly enough that the fractional 
changes in pressure, energy density, and gravitational field 
induced by the rotation are all relatively small \cite{H67}.  If 
$M$ and $R$ represent the mass and radius, respectively, of the 
non-rotating configuration, and $\Omega_{\n}$ and $\Omega_{\p}$ 
the respective constant angular speeds of the neutrons and protons, 
then ``slow'' can be defined to mean rotation rates that satisfy 
the inequality 
\beq
    \Omega_{\n}^2~{\rm or}~\Omega_{\p}^2~{\rm or}~\Omega_{\n} 
    \Omega_{\p} \ll \left({c \over R}\right)^2 {G M \over R c^2} 
    \ . \label{ineq}
\eeq 
Since $G M/R c^2 < 1$, the inequality also implies $\Omega_{\n,\p} 
R \ll c$, i.e.~that the linear speed of the matter must be much 
less than the speed of light.  

This is not as restrictive as one might guess, especially for the 
astrophysical scenarios we have in mind.  For a solar mass neutron 
star of radius 10~km, the square root of the combination on the 
right-hand-side of Eq.~(\ref{ineq}) works out to $11~500~{\rm 
s}^{- 1}$.  The Kepler limit (i.e.~the rotation rate at which 
mass-shedding sets in at the equator) for the same star is roughly 
$7700~{\rm s}^{-1}$.  The fastest known pulsar rotates with a 
period of $1.56~{\rm ms}$, which translates into a rotation rate of 
$4000~{\rm s}^{-1}$ or nearly half the Kepler limit.  This is why 
the slow-rotation approximation is still accurate to (say) $15-20~\%$ 
for stars rotating at the Kepler limit.  This is seen in 
Fig.~\ref{lorene}, which is taken from Prix et al \cite{PCA02}.  It 
shows a one-fluid rotating star's equatorial ($R_{\rm equ}$) and 
polar ($R_{\rm pole}$) radii, determined using both the slow-rotation 
approximation and the very accurate LORENE code developed by the 
Meudon Numerical Relativity group \cite{BGSM93,GHLPBM99}.  Notice 
that the slow-rotation approximation works quite well up to and 
including rotation rates equal to the fastest known pulsar.

The slow rotation scheme is based upon an expansion in terms of 
the constant rotation rates of the fluids.  The first rotationally 
induced effect that one encounters in general relativity is the 
linear order frame-dragging (i.e.~local inertial frames close to 
and inside the star are rotating with respect to inertial frames 
at infinity).  It is only at the second order that rotationally 
induced changes in the total mass, shape, and distribution of the 
matter of the star are produced.  Since the frame-dragging is a 
purely general relativistic effect, the Newtonian slow rotation 
scheme yields nothing at linear order, but does have second order 
changes that parallel those of general relativity.

Andersson and Comer \cite{AC01a} apply their slow rotation scheme 
using the two-polytrope equation of state with the model~III 
parameters listed in Table~\ref{modtab}.  Although their formalism 
is general enough to allow for entrainment, they do not consider 
it in their numerical solutions to the field equations, since their 
main emphasis is to extract effects due to two rotation rates that 
can be independently specified.  For instance, reproduced here in 
Fig.~\ref{frame1} are some of their typical results for the 
frame-dragging for modest values of the relative rotation 
$\Omega_{\n}/\Omega_{\p}$ between the neutrons and protons.  The 
solutions exhibit a characteristic monotonic decrease of the 
frame-dragging from the center to the surface of the star.  Although 
likely very unrealistic, they also considered an extreme case of 
having the neutrons and protons counterrotate, with the result shown 
in Fig.~\ref{frame2} (reproduced, again, from \cite{AC01a}).  Clearly 
the frame-dragging is no longer monotonic, and even changes sign.  
One can understand this behavior as follows: In the inner core of 
the star, the angular momentum in the protons dominates so that 
the frame-dragging is positive.  But in the outer layers of the 
star the fact that the neutrons contain roughly $90\%$ of the mass 
means they begin to dominate so that the frame-dragging reverses.  
Finally, we reproduce here one other result of Andersson and Comer 
in Fig.~\ref{kepom}, which is the effect of relative rotation on 
the Kepler limit.  When the relative rotation is larger than one 
there is little change in the Kepler limit, which is due simply to 
the fact that the neutrons contain most of the mass.  On the other 
hand, when the relative rotation is being decreased toward zero, 
the Kepler limit rises because the frequency of a particle orbiting 
at the equator is approaching the non-rotating limit (again because 
the neutrons carry most of the mass).      

Prix et al \cite{PCA02} apply their Newtonian slow rotation 
formalism in a manner similar to Andersson and Comer.  There 
are important differences, however, even beyond 
the exclusion of general relativistic effects, and these are 
(i) they use an equation of state that includes entrainment 
and terms related to symmetry energy (i.e.~terms 
\cite{PAL88} which tend to force the system to have as many 
neutrons as protons), and (ii) an exact solution to the slow 
rotation equations is used for the analysis.  Thus, they are 
able to explore how entrainment and the symmetry energy 
affects the rotational configuration of the star.  We reproduce 
here in Fig.~\ref{keplerlim} their result for the Kepler limit 
as the relative rotation is varied, for different values of 
entrainment (denoted $\varepsilon$) and symmetry energy 
(denoted as $\sigma$).  The big surprise is that the symmetry 
energy has as much an impact as the entrainment.  This fact has 
not been noticed before and should be explored in more depth, 
using a more realistic equation of state (like \cite{PAL88}). 

\section{The Linearized Non-radial Oscillations}

The second application of our superfluid formalism is to 
the problem of non-radial oscillations.  The ultimate goal 
is to calculate such oscillations, and the gravitational 
waves that result from them, for rotating neutron stars in 
the general relativistic regime.  This is not an easy task, 
and the problem has not been solved fully (for rapidly 
rotating backgrounds) even for the ``simpler'' case of the 
ordinary perfect fluid.  That is, while some recent progress 
has been made to calculate the frequency of oscillations 
\cite{Fetal02}, there are as yet no complete determinations 
of the damping rates of the modes because of gravitational 
wave emission.  Even using the slow rotation approximation 
there are complications, due to questions about the basic 
nature of the modes and if they can be separated into purely 
polar and axial parts, or if they are of the inertial hybrid 
mode class (as in \cite{LF99,LAF00}).  Nevertheless, we can 
gain valuable insight by considering non-radial oscillations 
on non-rotating backgrounds.  We will use the Newtonian 
equations to reveal the nature of the various modes of 
oscillation, by given the highlights of the recent analysis 
of Andersson and Comer \cite{AC01b}, and we summarize the 
main results of general relativistic calculations 
\cite{CLL99,ACL02} of mode frequencies and damping rates.  

\subsection{Linearized oscillations in Newtonian theory}

The investigation of the nature of the modes of oscillation 
in both the Newtonian and general relativistic regimes has 
now a decade and a half of history.  Epstein's work \cite{E88} 
is the beginning, since he is the first to suggest that there 
should be new oscillation modes because superfluidity allows 
the neutrons to move independently of the protons, and thereby 
increases the fluid degrees of freedom.  Mendell \cite{M91} 
reaches the same conclusion and moreover argues, using an 
analogy with coupled pendulums, that the new modes should have 
the characteristic feature of a counter-motion between the 
neutrons and protons, i.e.~in the radial direction as the 
neutrons are moving out, say, the protons will be moving in, 
which is to be contrasted with the ordinary fluid modes that 
have the neutrons and protons moving in more or less 
``lock-step''.  This basic picture has been confirmed by 
analytical and numerical studies 
\cite{LM94,L95,CLL99,SW00,AC01b,ACL02,PR02} and the new modes 
of oscillation are known as superfluid modes.  As we will see 
below they are predominately acoustic in nature, and have a 
sensitive dependence on entrainment parameters.  It is 
worthwhile to mention again that our equations are formally 
equal to the two-fluid set of Landau for superfluid He~II.  
With some hindsight, one recognizes that the superfluid modes 
could have perhaps been inferred to exist from a 
thermomechanical effect \cite{TT86} in which the normal fluid 
and superfluid are forced into counter-motion by passing an 
alternating current through a resistor placed in the container 
that holds the fluid. 

The existence of the superfluid modes seems to confirm one's 
intuition that a doubling of the fluid degrees of freedom should 
lead to a doubling of modes.  A review of the spectrum of modes 
for the ordinary fluid should then give one an idea of what to 
expect in the superfluid.  McDermott et al \cite{MHH88} have 
given an excellent discussion of many of the possible modes in 
neutron stars, and these include the polar (or spheroidal) f-, p-, 
and g-modes and the axial (or toroidal) 
r-modes.  But a puzzling aspect in all of this is that Lee's 
\cite{L95} numerical analysis does not reveal a new set of g-modes, 
and in fact does not find {\em any} g-modes of non-zero frequency.  
Of course, the model he considers is that of a zero-temperature 
neutron star, and so one would not really expect g-modes like 
those of the sun (which exist because of an entropy gradient) to 
be important in a mature neutron star.  However, Reisenegger and 
Goldreich \cite{RG92} have shown conclusively that a composition 
gradient, such as the proton fraction in a neutron star, will 
also lead to g-modes, and the model of Lee does have a composition 
gradient.  Fortunately, this issue has been clarified by 
Andersson and Comer \cite{AC01b} who use a local analysis of the 
Newtonian equations to confirm Lee's numerical result that there 
are no g-modes of non-zero frequency.  Their analysis of the 
zero-frequency subspace reveals two sets of degenerate spheroidal 
modes, which they take to be the missing g-modes.  They also 
find two sets of degenerate toroidal modes, which they interpret 
to be r-modes.  And in fact when they add in rotation they find 
that the degeneracy is lifted and two sets of non-zero frequency 
r-modes exist. 

Apart from questions about the g-modes, Andersson and Comer 
\cite{AC01b} also illuminate the character of the superfluid modes.  
Although they do not solve the linearized equations for global 
mode frequencies, they are able to demonstrate that the equation 
that describes the radial behaviour of the superfluid modes is of 
the Sturm-Liouville form for large frequencies $\omega$.  Thus, one 
can expect there to be a set of modes for which $\omega^2_n \to 
\infty$ as the index $n \to \infty$.  The equation that describes 
the ordinary fluid modes is also of the Sturm-Liouville form for 
large frequencies and so it also has a set of modes with the same 
mode frequency behavior.  That is both sets of modes will be 
interlaced in the pulsation spectrum of the neutron star.  

Finally, Andersson and Comer \cite{AC01b} also use a local analysis 
of the Newtonian equations to find a (local) dispersion relation for 
the mode frequencies.  Letting 
\beq
    c^2_\n \equiv {\n_{\n} \over m_{\n}} {\partial {\mu}_{\n} \over 
                    \partial \n_{\n}} \quad , \quad           
    c^2_\p \equiv {\n_{\p} \over m_{\p}} {\partial {\mu}_{\p} \over 
                \partial \n_{\p}} \ , 
\eeq 
and assuming that the proton fraction (i.e.~the ratio of the proton 
number density over the total number density) is small, then they 
find that one solution to the dispersion relation is  
\beq
    \omega^2_o \approx L_l^2,
\eeq
where 
\beq
    L_l^2 \approx {l (l + 1) c_{\n}^2 \over r^2} \ .
\eeq
Here $c_\n$ is essentially the speed of sound in the neutron fluid, 
$l$ is the index of the associated spherical harmonic $Y_l^m$ of 
the mode, and $r$ is the radial distance from the center of the star,   
and so this is the classic ordinary fluid solution in terms of the 
Lamb frequency $L_l$ \cite{UOAS89}.  Likewise, they find another 
solution of the form
\beq
    \omega^2_s \approx {m_\p \over m^*_p}~{l (l + 1) \over 
                       r^2}~c^2_\p \ ,
\eeq
where $c_\p$ is roughly the speed of sound in the proton fluid.  
Thus both solutions are of predominately acoustic nature, but we 
see that the second solution, which corresponds to the superfluid 
mode, has a sensitive dependence on entrainment (through the 
appearance of the proton effective mass).  An observational 
determination of the superfluid mode frequency via gravitational 
waves, say, could be used to constrain the proton effective mass 
\cite{AC01c,ACL02}.  This would translate into a deeper 
understanding of superfluidity at supra-nuclear densities since 
the effective proton mass is part of the input information for BCS 
gap calculations \cite{KKC01}.  

\subsection{Quasinormal modes in general relativity} \label{nonrad}

We have just seen that the spectrum of mode pulsations in a 
superfluid neutron star is significantly different from its ordinary 
fluid counterpart.  Moreover, we have also seen that the superfluid 
modes have a sensitive dependence on entrainment.  We will now 
confirm, and build on, this basic picture in a quantitative way by 
looking at numerical results for the modes obtained using the general 
relativistic formalism.  The key results to be described come from 
the work of Comer et al \cite{CLL99} and Andersson et al \cite{ACL02}.  
It is worth noting that the set of linearized equations that describe 
the mode oscillations of superfluid neutron stars has much in common 
with the ordinary fluid set, and so many of the computational and 
numerical techniques that have been developed for the ordinary fluid 
\cite{RW57,C64,TC67,PT69,T69a,T69b,CT70,LD83,DL85} can be adapted to 
the superfluid case.  That being said, Andersson et al \cite{ACL02} 
have developed a new technique for calculating the damping rates of 
the modes due to gravitational wave emission. 

Before we get to the ordinary fluid and superfluid modes, there is 
another set of modes, called w-modes, that we will discuss that 
exist only in a general relativistic setting.  They were first 
discovered by Kokkotas and Schutz \cite{KS92} and are due mostly to 
oscillations of spacetime, coupling only very weakly to the matter.  
For instance, Andersson et al \cite{AKS96} use an Inverse Cowling 
Approximation where all the fluid degrees of freedom are frozen out 
and were able to find the w-modes.  Comer et al \cite{CLL99} have 
obtained w-modes in the superfluid neutron star case and find that 
they look very much like those of ordinary fluid neutron stars.  
As well they do not find a second set of w-modes because of 
superfluidity.  Both results are due to the fact that the w-modes 
are primarily oscillations of spacetime. 

From the previous subsection we expect that there will be no g-modes 
in the pulsation spectrum (cf.~Sec.~\ref{zerofreq} where we find two 
sets of polar perturbations in the zero frequency subspace), but 
there should be interlaced in it the ordinary and superfluid modes.  
In general relativity one obtains quasinormal modes because each 
frequency will have an imaginary part due to dissipation via 
gravitational wave emission.  Such modes correspond to those 
particular solutions that have no incoming gravitational waves at 
infinity.  In Fig.~\ref{spectrum}, taken from \cite{ACL02}, is 
graphed the asymptotic amplitude of the incoming wave versus the real 
part of the mode frequency for model~I of Table~\ref{modtab}.  The 
zeroes of the asymptotic incoming wave amplitude correspond with the 
deep minima of the figure.  As expected, the ordinary and superfluid 
modes are interlaced in the spectrum and the lowest few have been 
identified in the figure.  

We recall that the ordinary fluid modes should be characterized 
by the neutrons and protons flowing in ``lock-step'' whereas the 
superfluid modes should have the particles in counter-motion.  As 
well, the superfluid modes should have a sensitive dependence on 
entrainment.  Both are confirmed by the next two figures  
(Figs.~\ref{entrain} and \ref{cross} both taken from \cite{ACL02}), 
which also reveal a phenomenon known as avoided crossings.  
Fig.~\ref{entrain} graphs the real part of the first few ordinary 
and superfluid mode frequencies as a function of the entrainment 
parameter $\epsilon$ (for the ``physical'' range discussed earlier 
in Sec.~\ref{sfform}).  The solid lines in the figure are for the 
ordinary fluid modes, and we see that the first few are essentially 
flat as the entrainment parameter is varied, but the superfluid 
modes (the dashed lines) are clearly dependent on the entrainment 
parameter.  Fig.~\ref{cross} is a graph of the Lagrangian variations 
in the neutron and proton number densities.  The two plots on the 
far left of the figure are for the modes whose frequencies are 
labelled by $a_0$ (for the superfluid mode) and $b_0$ (for the 
ordinary fluid mode) in Fig.~\ref{entrain}.  The $a_0$ graph in 
Fig.~\ref{cross} clearly indicates a counter-motion of the neutrons 
with respect to the protons, whereas the $b_0$ graph shows the 
opposite behavior.  

Another obvious feature of both figures is the avoided crossings 
phenomenon.  Near the top of Fig.~\ref{entrain} we see that there 
are points in the $(\epsilon,{\rm Re}~\omega M)$ plane where the 
solid and dashed lines approach each other, but just before 
crossing they diverge away from each other.  The most interesting 
aspect of an avoided crossing is how the mode functions behave 
before, during, and after the avoided crossing.  This is shown in 
Fig.~\ref{cross}, where the middle and right-hand-side graphs are 
for the modes of Fig.~\ref{entrain} labelled by 
$\{a_{0.1},b_{0.1}\}$ and $\{a_{0.2},b_{0.2}\}$, respectively.  
In the middle graphs we see that the modes no longer have a clear 
distinction as to whether or not the neutrons and protons are 
flowing together or in counter-motion.  But in the 
$\{a_{0.2},b_{0.2}\}$ graphs of Fig.~\ref{entrain} we see that 
now it is the superfluid modes that have the particles flowing 
together and the ordinary fluid modes show the counter-motion.  
Although we will not go into details here, Andersson and Comer 
\cite{AC01b} and Andersson et al \cite{ACL02} have suggested that 
this may explain one of the interesting results of Lindblom and 
Mendell \cite{LM00} on the effect of mutual friction damping on 
the r-modes in superfluid neutron stars, and that is that mutual 
friction damping is negligible for the r-modes except for a small 
subset of the values of the entrainment parameter $\epsilon$.  
Mutual friction should be most effective when the neutrons and 
protons are in counter-motion, as in the superfluid modes, 
and what may be happening is that as the entrainment parameter is 
varied it is possible that they find modes beyond an avoided 
crossing where the ordinary fluid modes take on the characteristic 
of counter-motion. 

\subsection{Detectable Gravitational Wave Signals?} \label{gws}

While the study of modes in superfluid neutron stars is a fascinating 
and complex mathematical and theoretical physics problem, the real 
hope is that one can use the results to make scientific progress.  
This is why it is important to understand superfluid neutron star 
dynamics for realistic astrophysical scenarios, because one wants 
to know if there are detectable gravitational waves that carry 
imprints of superfluidity.  In other words, one wants to develop a 
gravitational wave asteroseismology as a probe of neutron star 
interiors, in much the same way that helioseismology is already a 
probe of the sun and asteroseismology is a probe of distant stars 
\cite{sun}.  This exciting possibility is being discussed 
\cite{AK96,AK98,BCCS98,KAA01,AC01c} in the literature and already 
some quantitative statements are in hand.  Unfortunately, estimates 
for LIGO II suggest that one needs modes of unrealistic amplitudes 
to ensure detection.  The best possibility is for detection of modes 
following neutron star formation from gravitational collapse, but 
even this has to be qualified \cite{AK96,AK98} because of low event 
rates and uncertainties in the energy that will get deposited in the 
oscillations.  However, there is no reason why one should only be 
pessimistic, since clearly gravitational wave detection will improve 
as more experience is gained, and new technology will also lead to 
improved sensitivities.  For instance, there is already the so-called 
EURO \cite{euro} detector being discussed, which is a configuration 
of several narrow-banded (cryogenic) detectors operating as a 
``xylophone'' that should allow high sensitivity at high frequencies.  

Andersson and Comer \cite{AC01c} and Andersson et al \cite{ACL02} have 
taken the estimated spectral noise density for such a configuration 
and have used it to determine the signal-to-noise for a detection of 
oscillation modes from superfluid neutron stars that have been excited 
during a glitch.  These estimates have been for two of the best studied 
glitching pulsars, which are the Crab and Vela pulsars.  One assumes 
that a typical gravitational wave signal from a mode takes the form of 
a damped sinusoidal, where the damping time is that of the mode itself 
\cite{AC01c,ACL02}.  The amplitude of the signal can be expressed in 
terms of the total energy radiated through the mode.  Andersson and 
Comer and Andersson et al assume that this total energy is comparable 
to the amount of energy released during a glitch, which for the Crab 
and Vela pulsars can be of order $10^{- 12}-10^{- 13} M_{\odot} c^2$ 
\cite{FLE00,HHN00}.  Unfortunately, an error in the signal-to-noise 
calculation of Andersson and Comer, to be corrected in Andersson et al, 
has incorrectly estimated the predicted signal-to-noise for the EURO 
configuration.  Fortunately, the revised data still indicate sufficient 
sensitivity to expect detection for Crab- and Vela-like glitches.

\section{The general relativistic zero-frequency subspace} 
\label{zerofreq}

Andersson and Comer \cite{AC01b} have determined that the 
zero-frequency subspace in Newtonian theory is spanned by two sets 
of polar and two sets of axial degenerate perturbations.  These 
solutions are time-independent convective currents.  They were stated 
to be the two missing sets of g-modes and the two sets of r-modes that 
become non-degenerate when rotation is added.  We will extend this 
analysis to the general relativistic case and show that there are two 
sets of polar perturbations, thus supporting our earlier claim that 
there will be no non-zero frequency g-modes in the pulsation spectrum 
of a general relativistic superfluid neutron star.  We will also 
find two sets of axial perturbations, which could presumably lead to 
two sets of r-modes (or more general hybrid modes) when the background 
rotates.  

\subsection{The background spacetime and fluid configuration}

The background is treated exactly as in Comer et al. \cite{CLL99}, 
i.e.~it is spherically symmetric and static, and thus the metric 
can be written in the Schwarzschild form
\beq
  {\rm d}s^2 = - e^{\nu(r)} {\rm d}t^2 + e^{\lambda(r)} {\rm d}
               r^2 + r^2 \left({\rm d}\theta^2 + {\rm sin}^2
               \theta {\rm d}\phi^2\right) \ . \label{bgmet}
\eeq
The two conserved currents $\n^{\nu}$ and $\p^{\nu}$ are parallel 
with the timelike Killing vector $t^{\nu} = (1,0,0,0)$, and are 
thus of the form
\beq
    \quad \n^{\nu} = \n(r) {\cal U}^{\nu} \quad ,\quad \p^{\nu} = 
                     \p(r) {\cal U}^{\nu} \ , 
\eeq
where ${\cal U}^{\nu} = t^{\nu}/|{\bf t}|$.  Likewise, the 
chemical potential covectors $\mu_\nu$ and $\chi_\nu$ become  
\beq
    \chi_{\nu} = \chi(r) {\cal U}_{\nu} \ , \quad 
    \mu_{\nu} = \mu(r) {\cal U}_{\nu} \ ,
\eeq
where $\mu = \B \n + \A \p$ and $\chi = \C \p + \A \n$.  Explicit 
solutions for the background configurations can be constructed 
following the procedure of Comer et al \cite{CLL99}.  

\subsection{The linearized fluid and metric variables}

Making no assumptions yet on the metric and matter variations, 
we will first insert the background metric and matter variables  
into Eqs.~(\ref{dens_perbs}), (\ref{vel_perbs}) and 
(\ref{mom_perbs}).  Without too much effort, one finds 
\begin{eqnarray}
    \delta u^0 &=& {1 \over 2} e^{- 3\nu/2} \delta g_{00} 
                   \quad , \quad 
    \delta u^i = e^{- \nu/2} {\partial \over \partial t} 
                 \xi^i_\n \ , \cr
                && \cr
    \delta v^0 &=& {1 \over 2} e^{- 3\nu/2} \delta g_{00} 
                   \quad , \quad 
    \delta v^i = e^{- \nu/2} {\partial \over \partial t} 
                 \xi^i_\p \ ,
\end{eqnarray}
for the fluid velocity perturbations, 
\begin{eqnarray}
   \delta \n &=& - {\n \over 2} \left(e^{- \lambda} \delta g_{rr} 
                  + {1 \over r^2} \left[\delta g_{\theta\theta} + 
                  {1 \over {\rm sin}^2\theta} \delta g_{\phi\phi}
                  \right]\right) - {1 \over r^2 e^{\lambda/2}} 
                  {\partial \over \partial r} \left(\n r^2 
                  e^{\lambda/2} \xi^r_\n\right) - \cr
               && \cr
               && \n \left({\partial \over \partial \theta} 
                  \xi^\theta_\n + {\partial \over \partial \phi} 
                  \xi^\phi_\n + {\rm cot} \theta \xi^\theta_\n
                  \right) \ , \cr
               && \cr
   \delta \p &=& - {\p \over 2} \left(e^{- \lambda} \delta g_{rr} 
                  + {1 \over r^2} \left[\delta g_{\theta\theta} +
                  {1 \over {\rm sin}^2\theta} \delta g_{\phi\phi}
                  \right]\right) - {1 \over r^2 e^{\lambda/2}} 
                  {\partial \over \partial r} \left(\p r^2 
                  e^{\lambda/2} \xi^r_\p\right) - \cr
               && \cr
               && \p \left({\partial \over \partial \theta} 
                  \xi^\theta_\p + {\partial \over \partial \phi} 
                  \xi^\phi_\p + {\rm cot} \theta \xi^\theta_\p
                  \right) \ , \label{npvar}
\end{eqnarray}
for the density variations and 
\begin{eqnarray}
    \delta \mu_0 &=& {1 \over 2} \mu e^{- \nu/2} \delta g_{00} 
                     - e^{\nu/2} \left(\b00 \delta \n + \a00 
                     \delta \p\right) \ , \cr
                  && \cr
    \delta \mu_i &=& \mu e^{- \nu/2} \delta g_{0 i} + 
                     e^{- \nu/2} g_{i j} \left(\B \n {\partial 
                     \over \partial t} \xi^j_\n + \A \p 
                     {\partial \over \partial t} \xi^j_\p \right) 
                     \ , \cr
                  && \cr
    \delta \chi_0 &=& {1 \over 2} \chi e^{- \nu/2} \delta g_{00} 
                     - e^{\nu/2} \left(\c00 \delta \p + \a00 
                     \delta \n\right) \ , \cr
                  && \cr
    \delta \chi_i &=& \chi e^{- \nu/2} \delta g_{0 i} + 
                     e^{- \nu/2} g_{i j} \left(\C \p {\partial 
                     \over \partial t} \xi^j_\p + \A \n 
                     {\partial \over \partial t} \xi^j_\n \right) 
                     \ , \label{mom_vars}
\end{eqnarray}
for the momentum covector variations, where $\a00$, $\b00$, and 
$\c00$ can be calculated from Eq.~(\ref{coefficients}). 
One can also show that 
\beq
    \delta \Lambda = - \mu \delta \n - \chi \delta \p 
                  \quad , \quad
    \delta \Psi = \left(\b00 \n + \a00 \p\right) \delta \n + 
                  \left(\c00 \p + \a00 \n\right) \delta \p \ .
\eeq
Since the four velocity perturbations must be time-independent we 
see that the Lagrangian displacements must take the form  
\beq
    \xi^i_\n = e^{\nu/2} t \delta u^i + \zeta^i_\n 
               \quad , \quad
    \xi^i_\p = e^{\nu/2} t \delta v^i + \zeta^i_\p \ ,
\eeq
where $\zeta^i_\n$ and $\zeta^i_\p$ are ``integration constants'' 
(i.e.~they are independent of time but depend on the spatial 
coordinates).  We will see below that their only role is that 
once they have been determined they specify $\delta \n$ and 
$\delta \p$ via Eq.~(\ref{npvar}).  

Although the Einstein equations must be analyzed using a
decomposition in terms of spherical harmonics for the perturbations, 
we can actually completely solve the linearized Euler equations, 
which take the form
\beq
    \partial_t \delta \mu_i = \partial_i \delta \mu_t 
               \quad , \quad 
    \partial_t \delta \chi_i = \partial_i \delta \chi_t  
               \ . 
\eeq
One can easily verify that the left-hand-sides of each equation is 
zero (by taking a time derivative of $\delta \mu_i$ and $\delta 
\chi_i$ using Eq.~(\ref{mom_vars})) and it thus follows that  
\beq
    \delta \mu_t = 0 \quad , \quad \delta \chi_t = 0 \ .
\eeq
From Eq.~(\ref{mom_vars}) we see that these last two equations can 
be used to find $\delta \n$ and $\delta \p$ in terms of 
$\delta g_{0 0}$.  Given that the conservation equations are satisfied 
automatically by the Lagrangian displacements, then all of the fluid 
equations have been solved.   

We have seen earlier that the question of chemical equilibrium is 
an important aspect of the types of perturbations that can be 
induced on a superfluid neutron star.  Langlois et al \cite{LSC98} 
have argued that the general condition for chemical equilibrium to 
exist between the two fluids is that 
\beq
    \beta \equiv v^\nu \left(\mu_\nu - \chi_\nu\right) = 0 \ .
\eeq
For perturbations that do not maintain chemical equilbrium then 
it is the case that $\delta \beta \neq 0$.  Using the variations 
above we find
\beq
    \delta \beta = \left(\a00 -\b00\right) \delta \n - \left(\a00 
                   - \c00\right) \delta \p \ .
\eeq
We note here that axial perturbations are such that $\delta \n  
= \delta \p = 0$ and thus it follows that $\delta \beta = 0$, 
that is axial perturbations on spherically symmetric and static 
backgrounds must necessarily maintain chemical equilibrium 
between the two fluids if the background fluids were in chemical 
equilibrium.

\subsection{The Zero-Frequency Subspace}  

We have exhausted the information that can be extracted from 
just using the background configuration in the formulas for the 
variations.  To finish mapping out the zero-frequency subspace 
we must examine the Einstein equations.  In doing this it will be 
very convenient to consider two types of perturbations and those 
are the polar (or spheroidal) and the axial (or toroidal) 
perturbations.  For the metric perturbations we will use the 
Regge-Wheeler gauge \cite{RW57}, in which the polar components 
of the metric perturbations can be written as
\begin{equation}
     \delta^{P} g_{\mu \nu} = 
            \left[\matrix{e^{\nu(r)} H_{0}(r)&H_{1}(r)&0&0 
            \cr H_{1}(r)&e^{\lambda(r)} H_{2}(r)&0&0 \cr 
            0&0&r^2 K(r)&0\cr 0&0&0&r^{2} {\rm sin}^2\theta K(r)}
            \right]  Y^m_l(\theta,\phi) \ , \label{lemet}
\end{equation}
and the axial components as  
\beq
    \delta^{A} g_{\mu \nu} = \left[\matrix{0&0&h_{0}(r) \left(
            {- 1 \over {\rm sin}\theta}\right) {\partial \over 
            \partial \phi}&h_{0}(r) {\rm sin}\theta{\partial 
            \over \partial \theta}\cr 
            0&0&h_{\rm 1}(r) \left({- 1 \over {\rm sin}\theta}
            \right) {\partial \over \partial \phi} &h_{\rm 1}(r) 
            {\rm sin} \theta {\partial \over \partial \theta} \cr 
            h_{0}(r) \left({- 1 \over {\rm sin}\theta}\right) 
            {\partial \over \partial \phi}& h_{\rm 1}(r) 
            \left({- 1 \over {\rm sin} \theta}\right) {\partial 
            \over \partial \phi}&0&0\cr h_{0}(r) {\rm sin} 
            \theta{\partial \over \partial \theta}&h_{\rm 1}(r) 
            {\rm sin}\theta{\partial \over \partial 
            \theta}&0&0}\right] Y^m_l(\theta,\phi) \ , \ 
            \label{lomet}
\eeq
where the $Y^m_l$ are the spherical harmonics.  We write the polar 
unit four-velocity perturbations as
\beq
   \delta^{P} u^{\mu} = e^{- \nu/2} \left[\matrix{{1 \over 2} H_0
                        \cr
                        {1 \over r} W_{\n}(r)\cr
                        {1 \over r^2} V_{\n}(r) {\partial \over 
                        \partial \theta}\cr
                        {1 \over r^2 {\rm sin}^2\theta} V_{\n}(r) 
                         {\partial \over \partial \phi}}\right] 
                         Y^m_l(\theta,\phi) \quad , \quad
   \delta^{P} v^{\mu} = e^{- \nu/2} \left[\matrix{{1 \over 2} H_0
                        \cr
                        {1 \over r} W_{\p}(r)\cr
                        {1 \over r^2} V_{\p}(r) {\partial \over 
                        \partial \theta}\cr
                        {1 \over r^2 {\rm sin}^2\theta} V_{\p}(r) 
                        {\partial \over \partial \phi}}\right] 
                        Y^m_l(\theta,\phi) \ ,
\eeq
and the axial perturbations as
\beq
    \delta^{A} u^{\mu} = {e^{- \nu/2} \over r^2 {\rm sin}\theta} 
                         \left[\matrix{0 \cr 0 \cr
                         - U_{\n}(r) {\partial \over \partial 
                         \phi}\cr U_{\n}(r) {\partial \over 
                         \partial \theta}}\right] 
                         Y^m_l(\theta,\phi) 
                         \quad , \quad
    \delta^{A} v^{\mu} = {e^{- \nu/2} \over r^2 {\rm sin}\theta} 
                         \left[\matrix{0 \cr 0 \cr
                         - U_{\p}(r) {\partial \over \partial 
                         \phi}\cr U_{\p}(r) {\partial \over 
                         \partial \theta}}\right] 
                         Y^m_l(\theta,\phi) \ .
\eeq

Although we have already seen that the particle number density 
perturbations can be solved in terms of $\delta g_{0 0}$, it is 
worthwhile to comment just a little more on their form.  The 
easy case is that of axial perturbations since for them $\delta^A 
g_{0 0} = 0$ and so $\delta^A \n$ and $\delta^A \p$ must both 
vanish for a generic master function.  The polar perturbations 
in the particle number densities are a bit more complicated to 
determine, but in the end take a simple form.  Since $\delta^P 
\n$ and $\delta^P \p$ must both be time independent, a setting 
of time derivatives of Eq.~(\ref{npvar}) to zero yields 
constraints on the velocity perturbation functions, i.e.
\beq
    l (l + 1) V_\n = {e^{- \lambda/2} \over \n} \left(r \n 
                     e^{\lambda/2} W_\n\right)^{\prime} 
                     \quad , \quad
    l (l + 1) V_\p = {e^{- \lambda/2} \over \p} \left(r \p 
                     e^{\lambda/2} W_\p\right)^{\prime} \ . 
                     \label{vrelw}
\eeq
But recall that the Lagrangian displacements still have the 
``integration constant'' terms.  However, using the same 
type of decomposition for $\zeta^i_{\n,\p}$ as used for 
$\delta^P u^i$ and $\delta^P v^i$ above, and in place of the 
$W_{\n,\p}(r)$ and $V_{\n,\p}(r)$ coefficients we substitute 
some new coefficients $A_{\n,\p}(r)$ and $B_{\n,\p}(r)$, say, 
then it follows that  
\beq
    \delta^{P} \n = \delta \n(r) Y^m_l \quad , \quad 
    \delta^{P} \p = \delta \p(r) Y^m_l \ ,
\eeq
where the $\delta \n(r)$ and $\delta \p(r)$ are linear 
combinations of $A_{\n,\p}(r)$ and $B_{\n,\p}(r)$ and their 
derivatives.  These new coefficients $A_{\n,\p}(r)$ and 
$B_{\n,\p}(r)$ appear nowhere else but in $\delta^P \n$ and 
$\delta^P \p$ and thus this is why we stated earlier that the 
only role of the ``integration constants'' is to determine 
the particle number density perturbations.  Of course at this 
point we can forgo using $A_{\n,\p}(r)$ and $B_{\n,\p}(r)$ 
and just solve for the $\delta \n(r)$ and $\delta \p(r)$ 
instead.

After some algebra it can be shown that the perturbations 
for $l \geq 2$ result in three distinct groups of the linearized 
Einstein and superfluid field equations, and these are 

\bigskip
\noindent
(i) $l \geq 1$ Group I:
\begin{eqnarray}
    0 &=& e^{- \lambda} r^2 K^{\prime \prime} + e^{- \lambda} 
          \left(3 - {r \lambda^{\prime} \over 2}\right) r 
          K^{\prime} - \left({l (l + 1) \over 2} - 1\right) K - 
          e^{- \lambda} r H^{\prime}_0 - \left({l (l + 1) \over 
          2} - 1 - 8 \pi r^2 \Psi\right) H_0 \cr
       && \cr
       &&+ 4 \pi r^2 \left(\left[3 \chi - n \a00 - \p \c00\right] 
         \delta \p + \left[3 \mu - \p \a00 - \n \b00\right] 
         \delta \n\right) \ , \cr
     && \cr
    0 &=& e^{- \lambda} \left(1 + {r \nu^{\prime} \over 2}\right) 
          r K^{\prime} - \left({l (l + 1) \over 2} - 1\right) K - 
          e^{- \lambda} r H^{\prime}_0 + \left({l (l + 1) \over 
          2} - 1 - 8 \pi r^2 \Psi\right) H_0 \cr
       && \cr
       && + 4 \pi r^2 \left(\left[\chi - 3 \n \a00 - 3 \p \c00
          \right] \delta \p + \left[\mu - 3 \p \a00 - 3 \n \b00
          \right] \delta \n\right) \ , \cr
     && \cr
    0 &=& e^{- \lambda} r^2 K^{\prime \prime} + e^{- \lambda} 
          \left({r (\nu^{\prime} - \lambda^{\prime}) \over 2} + 
          2\right) r K^{\prime} - 16 \pi r^2 \left(\left[\n \a00 
          + \p \c00\right] \delta \p + \left[\p \a00 + \n \b00
          \right] \delta \n\right) \cr
       && \cr
       && - e^{- \lambda} r^2 H_0^{\prime \prime} - e^{- \lambda} 
          \left({r (3 \nu^{\prime} - \lambda^{\prime}) \over 2} + 
          2\right) r H_0^{\prime} - 16 \pi r^2 \Psi H_0\ , \cr
     && \cr    
    H_2 &=& H_0 \ , \cr
     && \cr
    K^{\prime} &=& e^{- \nu} \left(e^{\nu} H_0\right)^{\prime} 
                \ , \cr
                && \cr
    {\mu \over 2} H_0 &=& \a00 \delta \p + \b00 \delta \n \ , \cr
                       && \cr
    {\chi \over 2} H_0 &=& \a00 \delta \n + \c00 \delta \p \ .
\end{eqnarray}

\bigskip
\noindent
(ii) $l \geq 2$ Group II:
\bigskip
\begin{eqnarray}
    0 &=& H_1 + {16 \pi r e^{\lambda} \over l (l + 1)} \left(\mu 
          \n W_\n + \chi \p W_\p\right) \ , \cr
       && \cr
    0 &=& l (l + 1) V_\n - {e^{- \lambda/2} \over \n} \left(r \n 
                     e^{\lambda/2} W_\n\right)^{\prime} \ , \cr
       && \cr
    0 &=& l (l + 1) V_\p - {e^{- \lambda/2} \over \p} \left(r \p 
                     e^{\lambda/2} W_\p\right)^{\prime} \ , \cr
       && \cr
    0 &=& e^{- (\nu - \lambda)/2} \left(e^{(\nu - \lambda)/2} H_1
          \right)^{\prime} + 16 \pi e^{\lambda} \left(\mu \n V_\n 
          + \chi \p V_\p\right) \ .
\end{eqnarray}

\bigskip
\noindent
(iii) $l \geq 2$ Group III:
\bigskip
\begin{eqnarray}
   0 &=& h_0^{\prime \prime} - {\nu^{\prime} + \lambda^{\prime} 
         \over 2} h_0^{\prime} + \left({2 - l^2 - l \over r^2} 
         e^{\lambda} - {\nu^{\prime} + \lambda^{\prime} \over r} 
         - {2 \over r^2}\right) h_0 - 16 \pi e^{\lambda} 
         \left(\chi \p U_\p + \mu \n U_\n\right) 
         \ , \cr
      && \cr
   0 &=& (l - 1) (l + 2) h_1 \ , \cr
      && \cr
   0 &=& e^{- (\nu - \lambda)/2} \left(e^{(\nu - \lambda)/2} 
         h_1\right)^{\prime} \ .
\end{eqnarray}
A quick counting of the number of independent functions, and the 
number of equations, shows that Group I appears to have more 
equations than unknowns.  However, such a result is not 
unexpected because of the Bianchi identities.  For the present 
discussion, the important point about the Group I equations is 
that their solutions represent a subset that map static and 
spherically symmetric stars, with no mass currents, into other 
(nearby) static and spherically symmetric stars, also having no mass 
currents.  More interesting counting comes from the Group II and 
III equations.  For Group II one can show that the last equation 
in the group is a consequence of the other three.  Thus, we can 
specify arbitrarily $W_\n$ and $W_\p$, for instance, and then 
the other variables ($H_1$, $V_\n$, and $V_\p$) can be determined 
from the field equations.  Likewise, for Group III we can 
specify freely $U_\n$ and $U_\p$, and then the remaining variable 
$h_0$ is determined from its field equation (because it is clear 
that $h_1 =0$).

For various reasons, the cases of $l = 0,1$ must be distinguished 
from that of $l \geq 2$.  One reason is that there is more gauge 
freedom, which allows us to set $K(r) = h_1(r) = 0$ for $l = 1$ and 
in addition $H_1(r) = 0$ for $l = 0$ (which also has no axial 
perturbations).  Without listing all the formulas, we find that 
the counting of the number of equations and unknown functions is 
similar to $l \geq 2$.  In particular, the $l = 1$ analysis of the 
Groups II and III equations reveals that each have two arbitrary 
functions that must be specified before a solution can be 
obtained.  For $l = 0$, the polar currents must also vanish (otherwise 
they would diverge at the center of the star).  Hence, there are only 
$l = 0$ solutions that map static and spherically symmetric stars 
into other static and spherically symmetric stars.

\subsection{Decomposition of the zero-frequency subspace}

Regardless of the form of the equation of state for the 
background, or for the perturbations, we can make the following 
conclusions: Any solution 
\beq
    \{H_0,H_1,H_2,K,h_0,W_\n,W_\p,V_\n,V_\p,U_\n,U_\p,\delta 
    \n,\delta \p\}
\eeq
to the equations governing the time-independent perturbations of 
a static, spherical superfluid neutron star is a superposition 
of (i) a solution
\beq
    \{H_0,0,H_2,K,0,0,0,0,0,0,0,\delta \n,\delta \p\} 
\eeq
and (ii) a solution 
\beq
   \{0,H_1,0,0,h_0,W_\n,W_\p,V_\n,V_\p,U_\n,U_\p,0,0\} \ .
\eeq
The solutions in (i) are those that satisfy the Group I equations 
that map one static and spherically symmetric star to another (nearby)  
static and spherically symmetric star.  The solutions in (ii) are 
those that satisfy the Group II and III equations.  It is not 
difficult to see that the solutions (ii) contain two sub-classes, 
the purely polar solutions that satisfy the Group II 
equations, i.e.
\beq
   \{0,H_1,0,0,0,W_\n,W_\p,V_\n,V_\p,0,0,0,0\}
\eeq
and the purely axial solutions that satisfy the Group III 
equations, i.e.
\beq
   \{0,0,0,0,h_0,0,0,0,0,U_\n,U_\p,0,0\} \ .
\eeq
The first subclass is made of the g-modes because they are (1) 
purely polar and (2) the particle number densities (and likewise 
the energy density and pressure) vanish.  The second subclass is 
made of the r-modes because they are (1) purely axial and (2) the 
particle number densities (and likewise the energy density and 
pressure) vanish.  The final conclusion is that the zero-frequency 
subspace of superfluid neutron stars is spanned by two sets of 
g-modes and two sets of r-modes.  This is qualitatively the same 
conclusion as found for the Newtonian equations \cite{AC01a}, and 
we assert that there will be no non-zero frequency g-modes in the 
pulsation spectrum of non-rotating superfluid neutron stars.  
As for the axial perturbations, presumably rotation will break the 
degeneracy and we will find two sets of r-modes (or more general 
hybrid modes \cite{LF99,LAF00}), as in the Newtonian superfluid 
case, but this must be verified.  

\section{Concluding Remarks} \label{con}

We have reviewed recent work to model the rotation and 
oscillation dynamics of Newtonian and general relativistic 
superfluid neutron stars.  We have seen that superfluidity 
affects both the background and the perturbation spectrum 
of neutron stars and both should therefore cause an imprint 
of superfluidity to be placed in the star's gravitational 
waves.  In particular our local analysis of the Newtonian 
mode equations indicates that the superfluid modes should 
have a sensitive dependence on entrainment parameters, 
something that is supported by the quasinormal mode 
calculations.  Given a suitably advanced detector, like the 
EURO configuration, we have seen that gravitational 
waves emitted during a Vela glitch, say, should be detectable.  
The key conclusion is that direct detection of gravitational 
waves from glitching pulsars can be used to greatly improve 
our understanding of the local state of matter in superfluid 
neutron stars.  This may become more important in the next 
few years because of indications of free precession in neutron 
stars \cite{SLS00}, which if true means that the vortex creep 
model will have to be reconsidered.  

We have also put in place some groundwork for a 
future analysis of the CFS mechanism in superfluid neutron 
stars.  We have done this by using the ``pull-back'' formalism 
to motivate fluid variations in terms of constrained 
Lagrangian displacements.  We have also used them as 
perturbations to help map out the zero frequency subspace, and 
found that it is spanned by two sets of polar  
and two sets of axial perturbations.  It remains to be seen if 
adding rotation will lift the degeneracy and yield two sets of 
polar and two sets of axial oscillation modes, or if the more 
general inertial hybrid modes result.  

Finally, I would like to elaborate a little more on why I continue 
to treasure my two years in Israel with Jacob Bekenstein.  After I 
had completed working on the superfluid analogs of quantum field 
theory in curved spacetime effects I tried in vain to publish the 
work.  True to his character, Jacob had some very kind words of 
advise, which were to never worry that effort is lost when a project 
does not play out exactly as expected, because his own experience 
was that one, or many, pieces of it would eventually be of direct 
importance for something else.  For a young scientist, those were 
words of comfort and hope, and remain so for one that now has a 
little more experience.    

\acknowledgments

The bulk of the work discussed here is very much the result of a team 
effort, and would not have happened without my teammates Nils Andersson, 
David Langlois, Lap Ming Lin, and Reinhard Prix.  Thanks guys!  I also 
thank Ian Jones for providing me information and references about the 
estimates for galactic neutron star populations, and Brandon Carter and 
John Friedman for discussions on fluid variational principles.  Finally, 
I gratefully acknowledge partial support from a Saint Louis University 
SLU2000 Faculty Research Leave award, and EPSRC visitors grant 
GR/R52169/01 in the UK.

\vfill
\clearpage

\section*{Figure Captions}

\bigskip
\noindent
Figure~1.  The radial profiles of the neutron and proton background 
particle number densities, $n$ and $p$, respectively, for model~II. 
The model has been constructed such that it accords well with a 
$1.4M_\odot$ neutron star determined using the modern equation of 
state calculated by Akmal, Pandharipande and Ravenhall \cite{APR98}.  
For reference, we show as horizontal lines the number densities at 
which Akmal et al suggest that i) neutron drip occurs, ii) there is 
an equal number of nuclei and neutron gas, and iii) the crust/core
interface is located. It should be noted that the latter should not 
coincide with our core/envelope interface since one would expect 
there to be a region where crust nuclei are penetrated by a neutron 
superfluid.

\bigskip
\noindent
Figure~2.  Comparison of one-fluid slow-rotation configurations 
with numerical results obtained using the LORENE code. $R_{\rm 
equ}$ and $R_{\rm pole}$ are the star's equatorial and polar 
radius respectively. The stellar model is a $N = 1$ polytrope with 
mass \mbox{$M=1.4M_\odot$} and (static) radius $R = 10$~km.

\bigskip
\noindent
Figure~3.  Typical results for the frame-dragging $\omega$ for 
superfluid stars in which the neutrons and the protons rotate at 
slightly different rates.

\bigskip
\noindent
Figure~4.  The frame dragging in an extreme (rather unphysical) 
situation where the neutrons and the protons counterrotate in such 
a way that the net frame dragging is ``backwards'' at the surface 
of the star but ``forwards'' in the central parts.

\bigskip
\noindent
Figure~5.  The Kepler limit $\Omega_K$ is shown as a 
function of the relative rotation rate $\Omega_n/\Omega_p$
for our model star. The filled squares show the maximum 
allowed rotation rate for the protons ($\Omega_p$), while 
the open squares show the corresponding neutron spin rate
($\Omega_n$). The Kepler frequency simply corresponds to the 
largest of the two.

\bigskip
\noindent
Figure~6.  Plots of the neutron (\n) and proton (\p) Kepler 
limits as functions of the relative rotation 
$\Omega_\n/\Omega_\p$, for $\sigma = -0.5,0,0.5$ and 
$\varepsilon = 0,0.4,0.7$.

\bigskip
\noindent
Figure~7.  This figure shows the asymptotic amplitude $A_{\rm in}$ as
a function of the (real) frequency $\omega M$ for our model~I. 
The slowly-damped QNMs of the star show up as zeros of $A_{\rm in}$, 
i.e.~deep minima in the figure. The first few ``ordinary'' and 
``superfluid'' modes are identified in the figure.

\bigskip
\noindent
Figure~8.  This figure shows how the frequencies of the fluid 
pulsation modes for our model~II vary with the entrainment parameter 
$\epsilon$.  The modes shown as solid lines are such that the two 
fluids are essentially comoving in the $\epsilon \to 0$ limit, while 
the modes shown as dashed lines are countermoving. As is apparent from 
the data, the higher order modes exhibit avoided crossings as $\epsilon$ 
varies. Recall that the range often taken as ``physically relevant'' 
is $0.04 \le \epsilon \le 0.2$. We  indicate by $a_\epsilon$ and 
$b_\epsilon$ the particular modes for which the eigenfunctions are 
shown in Fig.~\ref{cross}.

\bigskip
\noindent
Figure~9.  An illustration of the fact that the modes exchange 
properties during an avoided crossing.  We  consider two modes, 
labelled by $a_\epsilon$ and $b_\epsilon$ (cf. Figure~\ref{entrain}).  
The mode eigenfunctions are represented by the two Lagrangian number 
density variations, $\Delta n$ and $\Delta p$ (solid and dashed 
lines, respectively).  For mode $a$ the two fluids are essentially 
countermoving in the $\epsilon\to 0$ limit (it is a superfluid mode), 
while the two fluids comove for mode $b$ (it is similar to a standard 
p-mode). After the avoided crossing (which takes place roughly 
at $\epsilon=0.1$) the two modes have exchanged properties.

\vfill
\clearpage

\section*{Table Captions}

\bigskip
\noindent
Table~1.  Parameters describing our stellar models~I, II and III. 
Model~I is identical to model~2 of \cite{CLL99}, and has only a 
core with no envelope.  Model~II has an envelope of roughly 1~km 
and could be seen as a slightly more realistic neutron star model.  
Model~III has no distinct envelope but does have a more realistic 
mass and radius than Model~I.

\begin{figure}
\centering
\caption{}
\vskip 24pt
\includegraphics[clip,width=14cm]{Lorene_compare.eps} 
\label{lorene}
\end{figure}   

\begin{figure}[h]
\centering
\caption{}
\vskip 24pt
\includegraphics[height=10cm,clip]{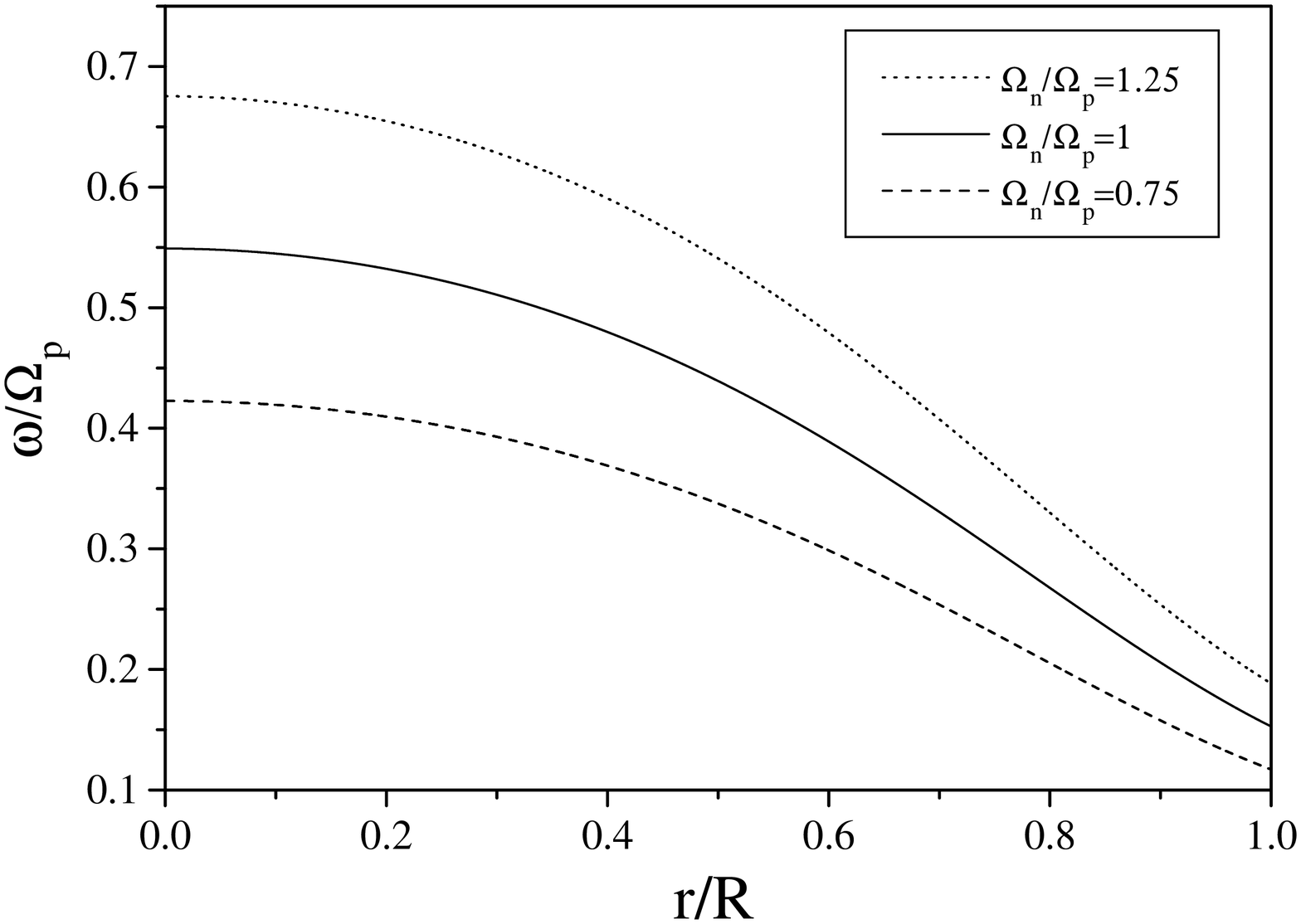}
\label{frame1}
\end{figure}

\begin{figure}[h]
\centering
\caption{}
\vskip 24pt
\includegraphics[height=10cm,clip]{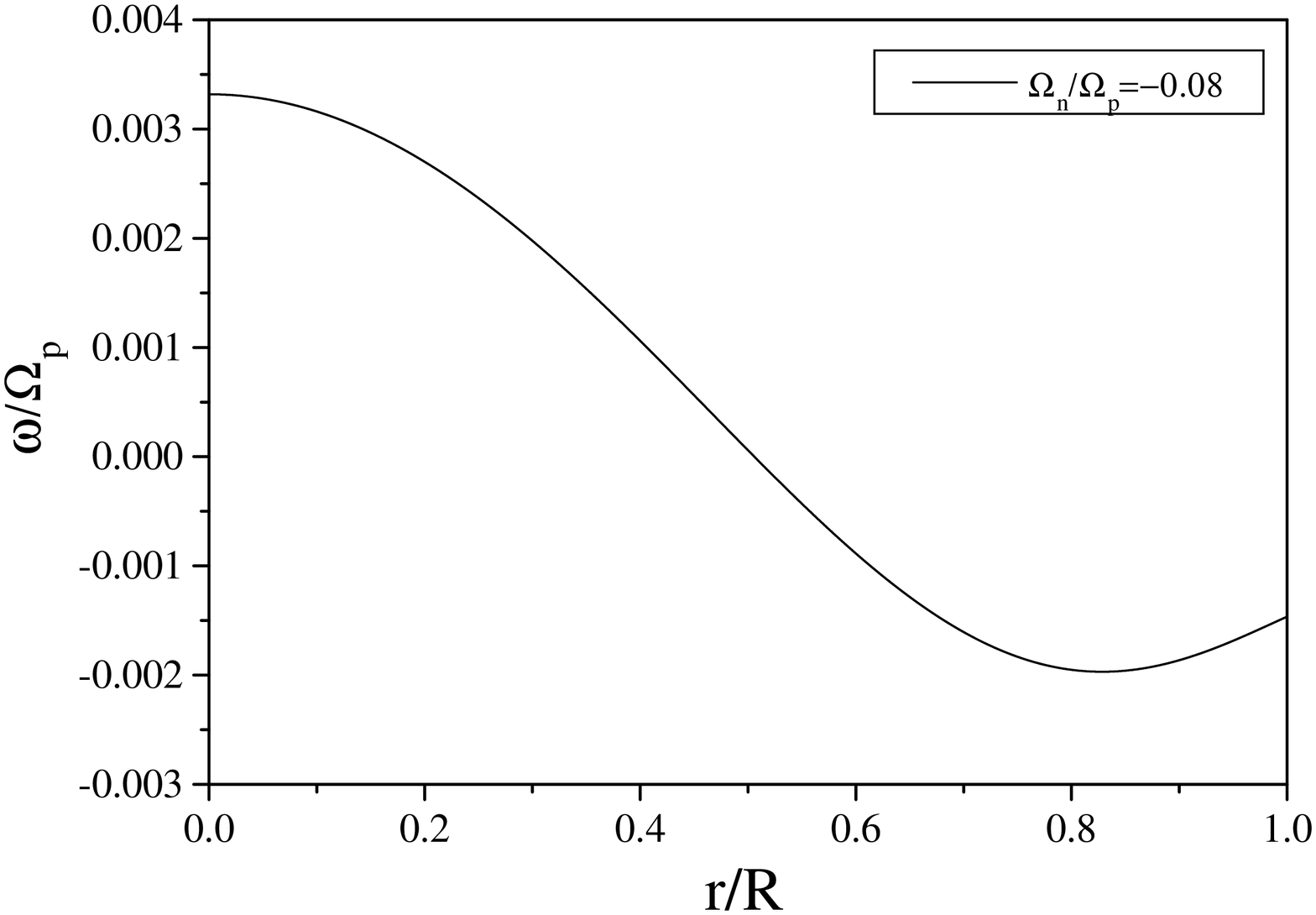} 
\label{frame2}
\end{figure}

\begin{figure}[h]
\centering
\caption{}
\vskip 24pt
\includegraphics[height=10cm,clip]{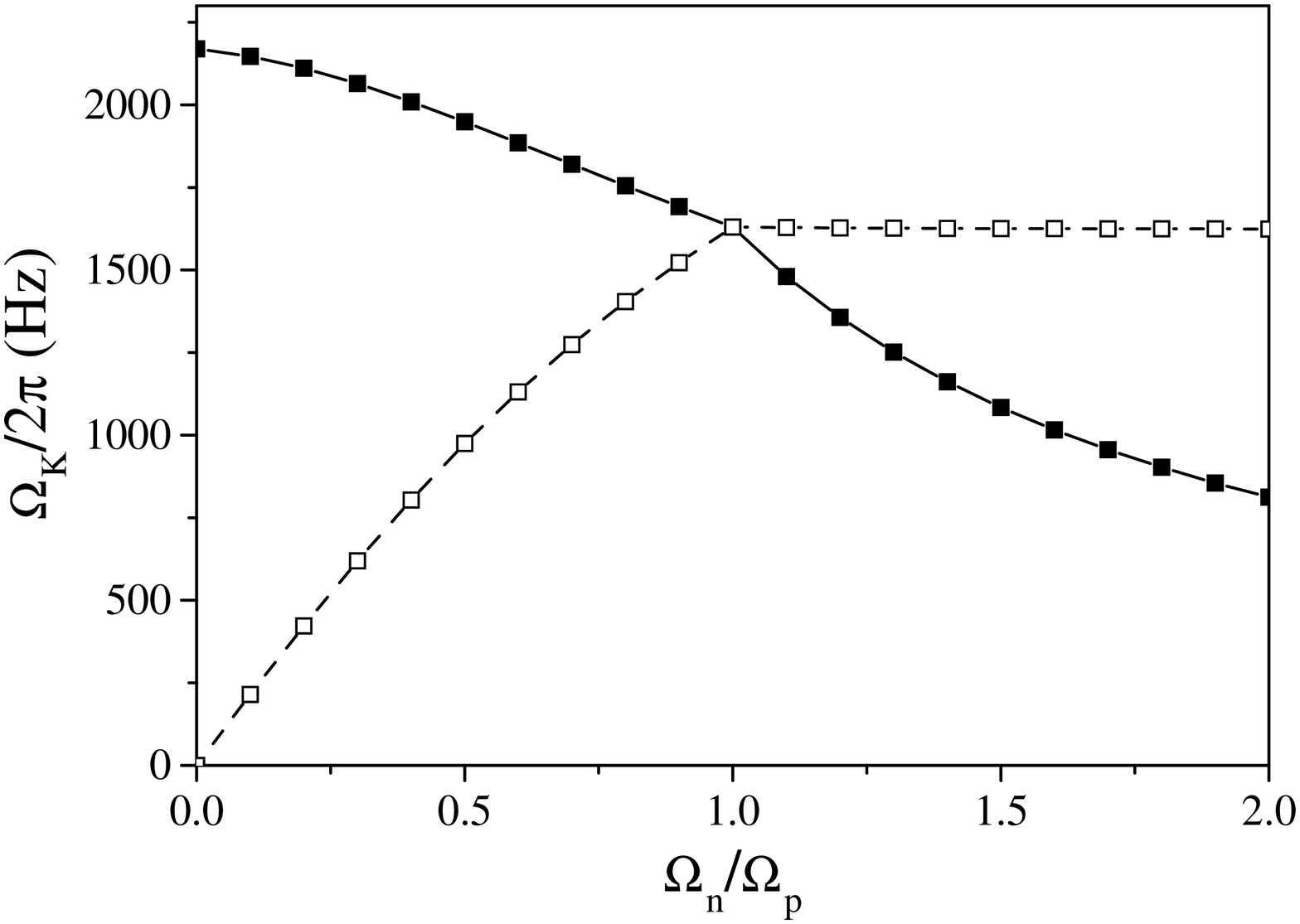}
\label{kepom}
\end{figure}

\begin{figure}
\centering
\caption{}
\vskip 24pt
\includegraphics[clip,width=16cm]{NewKepler.eps} 
\label{keplerlim}
\end{figure}

\begin{figure}[t]
\centering
\caption{}
\vskip 24pt
\includegraphics[height=10cm,clip]{Ainplot.eps}
\label{spectrum}
\end{figure}

\begin{figure}[h]
\centering
\caption{}
\vskip 24pt
\includegraphics[height=10cm,clip]{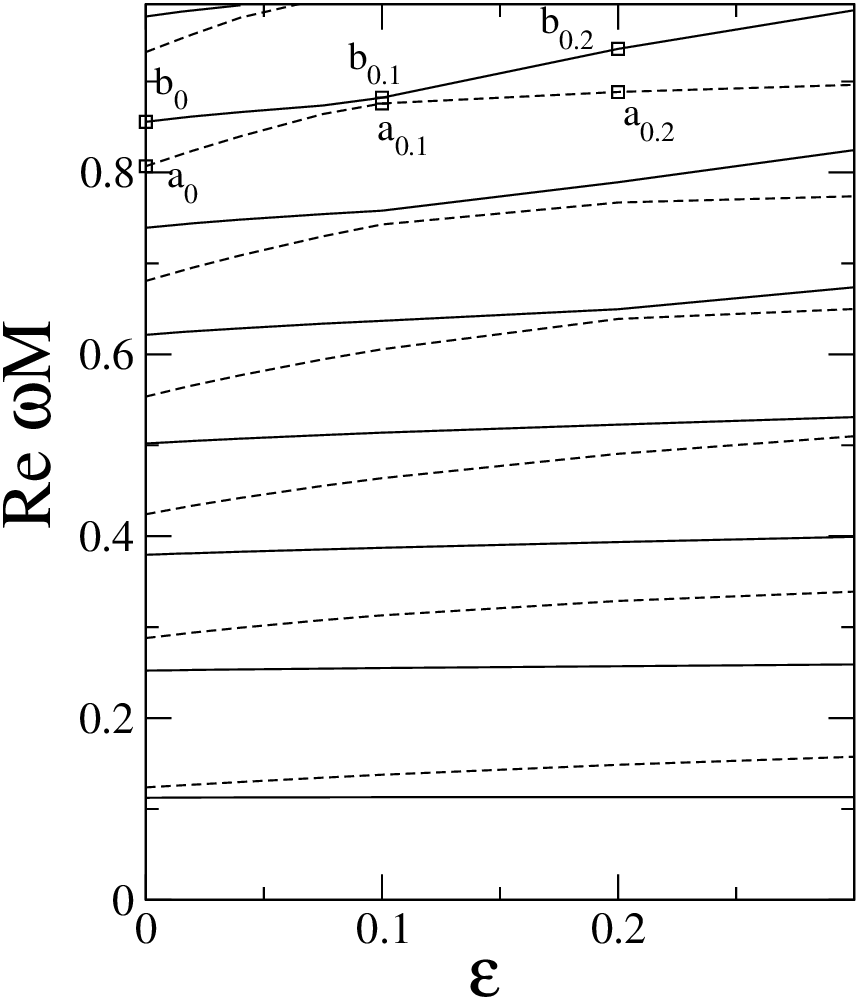}
\label{entrain}
\end{figure}

\begin{figure}[h]
\centering
\caption{}
\vskip 24pt
\includegraphics[height=10cm,clip]{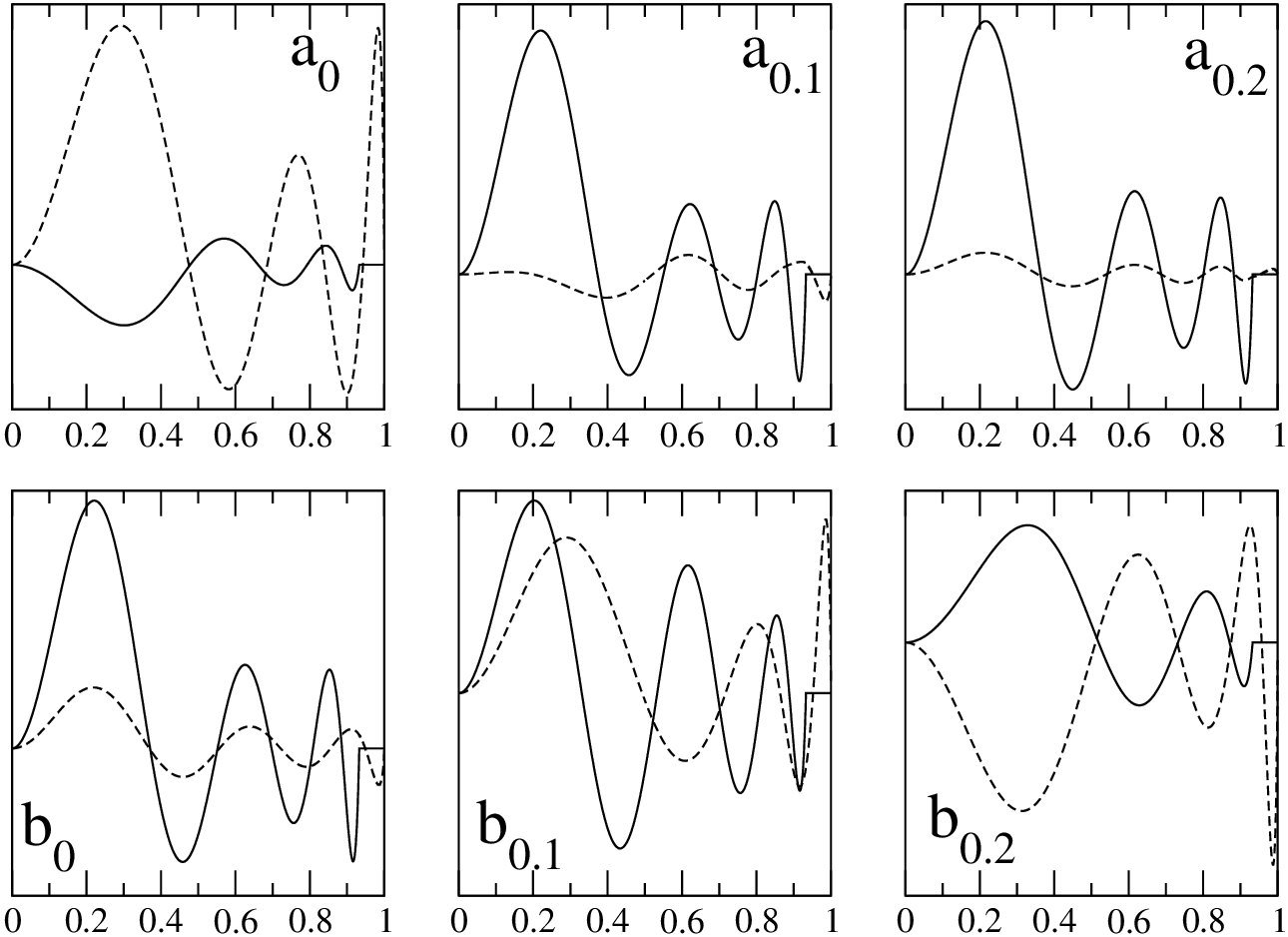} 
\label{cross}
\end{figure}

\begin{table}
\caption{}
\vskip 24pt
\begin{tabular}{|c|c|c|c|}
\hline
 & model I & model II & model III\\
\hline
$\sigma_n/m_n$ & 0.2 & 0.22 & 0.2 \\
$\sigma_p/m_n$ & 0.5 & 1.95 & 2 \\
$\beta_n$ & 2.5 & 2.01 & 2.3 \\
$\beta_p$ & 2.0 & 2.38 & 1.95 \\
$n_c$ (fm$^{-3}$) & 1.3 & 1.21 & 0.93 \\
$p_c$ (fm$^{-3}$) & 0.741 & 0.22 & 0.095 \\
$M/M_\odot$ & 1.355 & 1.37 & 1.409 \\
$R\ (\hbox{km})$ & 7.92 &  10.19 & 10.076 \\
$R_{c}\ (\hbox{km})$ & --- & 8.90 & --- \\
\hline
\end{tabular}
\label{modtab}
\end{table}


\begin{thebibliography}{10}
\bibitem{rationale}See \cite{CL93} for a complete discussion as 
to why three (and not four) scalar fields per four-current are 
all that is really needed to have a complete description of the 
dynamics.
\end{thebibliography}
\end{document}